\documentclass[10pt,journal,compsoc]{IEEEtran}
\usepackage[table]{xcolor}
\usepackage{graphics}
\usepackage{graphicx}
\usepackage[style=base]{caption}
\makeatletter
\g@addto@macro\normalsize{%
  \setlength\abovedisplayskip{40pt}
  \setlength\belowdisplayskip{40pt}
  \setlength\abovedisplayshortskip{40pt}
  \setlength\belowdisplayshortskip{40pt}
}

\makeatother

\usepackage[T1]{fontenc}
\usepackage[ruled]{algorithm2e}
\usepackage{float}
\usepackage[caption=false]{subfig}
\usepackage{enumitem}
\usepackage{array}
\usepackage{multirow}
\usepackage[symbol]{footmisc}
\usepackage{ragged2e}
\usepackage{verbatim}
\usepackage{enumitem}
\usepackage{amssymb}
\usepackage{booktabs}
\newtheorem{proposition}{Proposition}

\begin{document}
	
	\title{Application of Blockchain in Healthcare and Health Insurance Sector}
	
	\author{Debendranath Das, Indian Statistical Institute, debendra\_r@isical.ac.in, India}
	
	\IEEEtitleabstractindextext{%
		\begin{abstract}
		    \justify
			Technology has evolved over the years, making our lives easier. It has impacted the healthcare sector, increasing the average life expectancy of human beings. Still, there are gaps that remain unaddressed. There is a lack of transparency in the healthcare system, which results in inherent trust problems between patients and hospitals. In the present day, a patient does not know whether he or she will get the proper treatment from the hospital for the fee charged. A patient can claim reimbursement of the medical bill from any insurance company. However, today there is minimal scope for the Insurance Company to verify the validity of such bills or medical records. A patient can provide fake details to get financial benefits from the insurance company. Again, there are trust issues between the patient (i.e., the insurance claimer) and the insurance company. Blockchain integrated with the smart contract is a well-known disruptive technology that builds trust by providing transparency to the system. In this paper, we propose a blockchain-enabled \emph{Secure and Smart HealthCare System}. Fairness of all the entities: patient, hospital, or insurance company involved in the system is guaranteed with no one trusting each other. Privacy and security of patient's medical data are ensured as well. We also propose a method for privacy-preserving sharing of aggregated data with the research community for their own purpose. Shared data must not be personally identifiable, i.e, no one can link the acquired data to the identity of any patient or their medical history. We have implemented the prototype in the Ethereum platform and Ropsten test network, and have included the analysis as well.
		\end{abstract}
		
	}
	
	\maketitle
	
	\section{Introduction}
	The use of IoT devices in the healthcare sector generates a massive amount of patient's medical data \cite{reyna2018blockchain, dorri2017blockchain}. The EHRs (Electronic Health Records) are typically stored in some databases. However, these data are sensitive and cannot be made public. Tampering of medical data has the potential to jeopardize a person's life. Access to medical records must not compromise the privacy of a person. A third party might use such sensitive data to inflict harm or sell it to other parties.  
	Blockchain is a distributed immutable verifiable append-only ledger, which is maintained by a peer-to-peer network \cite{nakamoto2008re, zheng2017overview}. A block in a blockchain contains a set of transactions or data. Every block in the chain is connected to the previous one by a secure hash function, ensuring immutability. Immutability guarantees that once the transaction or data is written in some block in the blockchain, it cannot be tampered. Another essential property of Blockchain is its decentralized and distributed nature. There is no central authority that can solely handle the data in the blockchain. In a permissionless Blockchain model, anyone can download and store the ledger locally and then take part in the consensus protocol to sync the state of their blockchain with others.
	The following salient features of blockchain make it an attractive technology for addressing the security and privacy issues in IoT based applications\cite{zheng2018blockchain}:
	\begin{itemize}[leftmargin=*]
		\item Decentralization: The lack of central authority ensures robustness and overcomes the problem of a single point of failure. Moreover, any participant can join the system and function without being controlled by any single party.
		\item Anonymity: The inherent anonymity is well-suited for most IoT use cases, where the identity of the users must be kept private.
		\item Security: Blockchain realizes a secure network over untrusted parties that are desirable in IoT with many heterogeneous devices. An immutable ledger also helps to achieve non-repudiation.
		\item Data immutability: Blockchain technology is expected to improve data record management and brings fairness and auditability in the system.
	\end{itemize}
	Besides this, smart contracts come into the picture with the advent of Ethereum\cite{buterin2014next}. Contracts are meant for agreement between two or more entities. Smart contracts introduce a set of advantages, such as cost reduction, speed, precision, efficiency, and transparency. In many applications, a smart contract helps to eliminate the role of a trusted third party and brings fairness to the system.
	
	We have designed a patient-centric healthcare system where the medical data can be accessed by a party provided the patient has given consent to do so. An access control matrix stored in Blockchain disallows any sort of malicious intervention. The digital footprint of the medical data is stored in the blockchain to provide integrity, immutability, and digital signature also helps to provide accountability as well. The hospital authority cannot extort any arbitrary amount from a patient by providing wrong treatment or denying treatment. In that case, the hospital authority will be penalized. On the other hand, if a patient claims that he/she has received an invalid medical report or denies receiving treatment, the logic encoded in the smart contract prevents such behavior by penalizing the patient.
	
	Transcript of any interaction between patient and insurance company or database owner prevents any party from wrongly ascribing blame to an honest party. Such audit trails help in resolving any dispute by seeking legal help or mutual settlement. 

	\subsection{Related Work}
    We discuss the state-of-the-art blockchain solution for the healthcare system. Xia et al. \cite{MedShare} had proposed a cloud-based blockchain platform for sharing files with untrustworthy parties seeking access to medical files. Accessing continuous stream of patient data was enabled using blockchain architecture in \cite{ContinuousPatientMonitoring}. To reduce the overhead of data stored in Blockchain, Zhang et al. \cite{FHIRChain} proposed a blockchain-based solution where data is stored offline and the verification part of the associated data is put on the blockchain.

    
    Griggs et al. \cite{RemotePatientMonitoring} proposed a solution where the data from wearables act as the oracle to the smart contracts in the blockchain. The data preservation system in the work proposed by Li et al. \cite{DataPreservationSystem} basically contains two programs - the data access program and the blockchain interaction program. 
    \newline
    Dias et al. \cite{AccessControl} proposed a blockchain-based access control method for EHR. The policy is a relation between an EHR and the associated data keepers. Liang et al. \cite{DecentralizedAccountability}  proposed a three-layer architecture for securely storing and retrieving data. The continuous data from the wearables are stored in the cloud and their corresponding hashes are kept in the blockchain. 
    \newline
    Yup et al. \cite{yue2016healthcare} designed the \emph{healthcare data gateway}, a blockchain approach for healthcare intelligence to address the privacy of users by proposing a data access control for privacy. Zhang et al. \cite{zhang2016secure} proposed PSN-based healthcare by designing two protocols for authentication and sharing of healthcare data. Liang et al. \cite{liang2017integrating} proposed a mobile-based healthcare record sharing system using blockchain. They designed a secure user-centric approach to provide access control and privacy using a channel formation scheme. Jiang et al. \cite{jiang2018blochie} designed a medical data exchange system using blockchain by developing off-chain and on-chain verification for the security of the system’s storage.
    \newline
    Fan et al. \cite{fan2018medblock} proposed an improved consensus mechanism to obtain enhanced security and privacy of medical data. Wang and Song \cite{wang2018secure} proposed an attribute-based encryption mechanism to design a secure medical record sharing system.
    \cite{guo2018secure} presented an electronic health record management system using an attribute-based signature scheme for multiple users in the blockchain network. Sun et al. \cite{sun2018decentralizing} designed a distributed attribute-based signature scheme for medical systems based on blockchain and proposed a blockchain-based record sharing protocol. 
    \newline
    Zhang and Poslad \cite{zhang2018blockchain} proposed an access control policy for electronic medical records with finer granular access. Yang and Li \cite{yang2018design} proposed an architecture for securing EHR based on distributed ledger technology. Performance of Hyperledger Fabric, a permissioned blockchain network, was analyzed in \cite{sukhwani2017performance}. Gorenflo et al. \cite{gorenflo2020fastfabric} proposed a performance optimization for Hyperledger blockchain framework.  A detailed analysis of these papers have been provided in \cite{TANWAR2020102407}.
    
    \begin{table}[!ht]
        \centering
        \caption{Drawbacks of State-of-the-Art Healthcare Systems}
        \label{tab: Shortcomings of Papers}
        \scalebox{0.8}{
            \begin{tabular}{p{0.35\linewidth} | p{0.6\linewidth}}
                \hline \hline
        		\textbf{Related Works} & \textbf{Drawbacks} \\ \hline \hline
                \cite{yue2016healthcare,liang2017integrating,zhang2018blockchain,yang2018design} & Lack implementation and do not evaluate System's Performance \\ \hline
                \cite{MedShare,zhang2016secure,liang2017integrating,zhang2018blockchain,yang2018design} & Lack Proper Framework \\ \hline
                \cite{DataPreservationSystem,zhang2016secure} & Do not provide Data Access Control Policy \\ \hline
                \cite{yue2016healthcare,liang2017integrating,zhang2018blockchain,yang2018design,sukhwani2017performance} & No Algorithm is given for the Protocol \\ \hline
                \cite{yue2016healthcare,zhang2016secure,jiang2018blochie,sun2018decentralizing,zhang2018blockchain,yang2018design,sukhwani2017performance,gorenflo2020fastfabric} & Lack EHR Sharing Protocol \\ \hline
                \cite{MedShare,sukhwani2017performance} & Scalability Problem \\ \hline
                \cite{liang2017integrating,guo2018secure} & Interoperability Problem \\ \hline
                \cite{MedShare} & Key Management Problem \\ \hline
                \cite{ContinuousPatientMonitoring,jiang2018blochie,sun2018decentralizing} & Performance and Fairness problem \\ \hline
                \cite{fan2018medblock,sun2018decentralizing,gorenflo2020fastfabric} & Require High Storage, Power and/or Computation Cost\\ \hline 
            \end{tabular}
        }
    \end{table}
    
    Although state-of-the-art tries to enhance the security of the healthcare system using Blockchain Framework, these have certain drawbacks. We have summarized it in Table~\ref{tab: Shortcomings of Papers}. In our model, we have addressed most of the existing problems by proposing a decentralized, distributed healthcare system using a permissionless blockchain framework. We have implemented our protocol and evaluated the system's performance. Our model also ensures the privacy of patients' medical data (using Access Control Policy), data security, and fairness of various entities involved in the system.

	\subsection{Contribution of the Paper}
	In this paper, we define and discuss in detail a new blockchain-based architecture for a patient-centric Healthcare system. Our system is robust, but at the same time privacy-preserving. We briefly discuss the novelty of the design:  
	\begin{itemize}[leftmargin=*]
	\item We propose a \emph{Secure and Smart HealthCare System} which coordinates the interaction between patient and hospital while a patient is getting treated; between patient, the medical database owner, and insurance company when the patient needs to claim any reimbursement and between medical database owner and the research community while collecting the data.
	\item The proposed system ensures that no one has access to the patient's data stored in the medical database until and unless the patient has granted permission to do so.
	\item A patient cannot be overcharged for seeking treatment from a hospital. Simultaneously, a hospital has to start the treatment within a specified period of time. If they fail to do so, the patient can withdraw any deposit made.
	\item A patient can either retain the data with himself or store it in a medical database, which is assumed to be trusted/semi-trusted. The database owner will impose a check on the validity of the data provided by the patient before storing it in the database that prevents storing any spurious data.
	\item Fairness is guaranteed for both the patient and the insurance company when the former places a request for a claim based on an insurance policy purchased. An insurance company can verify whether the claim is valid. 
	\item Patient's data remains private and gets shared among those who have been granted legitimate access. Data cannot be tampered by any external agent, any malicious behavior can be detected using the digital fingerprint of the data recorded in Blockchain.
	\item Medical data provided to the research community by the system for study or analysis will not compromise the patient's privacy. 
	
	\item We have implemented the prototype in the Ethereum platform and Ropsten test network, and have evaluated the performance. Code for the protocol is available on GitHub\footnote{\label{note1} https://github.com/Debendranath/Blockchain-Enabled-Secure-and-Smart-Healthcare-System}.

	\end{itemize}
	
	\subsection{Organization}
	The rest of the paper is structured as follows - \\
	Section 2 briefly discusses the basic building blocks. In Section
	3, the system model and high-level view of our construction are presented. We have discussed our major procedures, implementation, and technical details in Section 4. In Section 5, we have addressed our security claims. Section 6 presents the results of our proposed system. Finally, we have concluded the paper in Section 7.
	
	\section{Preliminaries}
	In this section, we discuss the building blocks and other proposed methodologies that we have used in our system.The set of all binary strings of length $n$ is denoted as $\{0,1\}^n$, and the set of all finite binary strings as $\{0,1\}^*$. The output \(x\) of an algorithm \( \mathcal{A} \) is denoted by \(x\) $\leftarrow$ \( \mathcal{A} \).
	\subsection{Basic Cryptographic Primitives}
\begin{figure}[!ht]
     		\begin{center}  
     			\includegraphics[width=3.5in, height=1.9in]{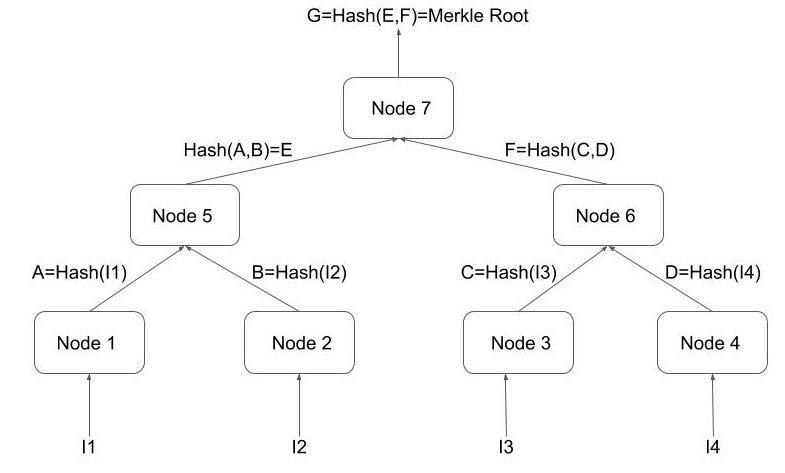}
     			\setlength{\belowcaptionskip}{-10pt}
     			\caption{\small \sl Merkle Tree
     				\label{fig:Image1}}
     		\end{center}  
 	\end{figure}

	\begin{itemize}[leftmargin=*]
		\item \textbf{Encryption Scheme:} Encryption Scheme serves the purpose of privacy or confidentiality of the messages exchanged between a sender and recipient. We have used two types of encryption - \textit{Private Key Encryption} or \textit{Symmetric Key Encryption \textbf{(SKE)}} and \textit{Public Key Encryption} or \textit{Asymmetric Key Encryption \textbf{(ASKE)}}.

		\item \textbf{Digital Signature Scheme:} Digital Signature (DS) serves the purpose of user authentication in the system. Obtaining the DS of an entity, any recipient can verify if the message originated from the intended sender or not.
		
		\item \textbf{Hash Function:} A Hash function ensures data integrity in the system. It can be defined as \textbf{$H:\{0,1\}^* \rightarrow \{0,1\}^k$} that maps messages of arbitrary length to a fixed size message digest of length $k$. An ideal hash function must satisfy the following three properties: \emph{One Way or Pre-image Resistant}, \emph{ Second Pre-image Resistant} and \emph{Collision Resistant}. 
	
	\item \textbf{Merkle Tree:} Merkle Trees are binary trees, which are used to prove the membership of data belonging to a set. The leaf node comprises the data present in the set, and the output of a leaf node is the hash of the data. For every non-leaf node, the output is the hash of the concatenated children node’s outputs. The output of the root node is referred to as the \emph{Merkle Root}. The acronym MR is used to denote the Merkle Tree Root, as shown in Figure~\ref{fig:Image1}.
	\end{itemize}
	
	\subsection{Other Building Blocks}
	
    \begin{itemize}[leftmargin=*]
    
        \item \textbf{Blockchain:} A Blockchain (BC) is an immutable, decentralized, distributed verifiable ledger. Once the information is put into the BC, it can not be tampered with by any means later. As the name suggests, it consists of a chain of blocks, where each block in turn consists of a series of transactions. The blocks are typically \emph{hash-linked} to one another, i.e, if a transaction or data inside any block is altered, the hash of that particular block gets changed. A mismatch in the hash value gets reflected in the succeeding blocks. Hence, to prevent such detection, the hash needs to be altered in subsequent blocks as well. In practice, it is not possible to do (or can be done with negligible probability), and that is the foundation of the immutability in Blockchain. The most commonly used consensus mechanism is Proof-of-Work (PoW) in Bitcoin as well as Ethereum. Other alternate consensus algorithms like Proof-of-Stake, Proof-of-Burn, Proof-of-Authority are still being explored. In our work, we use a Proof-of-Work (PoW) based consensus mechanism.
        
        \item \textbf{Smart Contract:} Smart Contracts (SC) are self-executing contracts, where the terms of an agreement between parties are encoded using a programming language(e.g, solidity). The smart contract also helps to build trust in the system by eliminating the role of intermediaries or middle man. Blockchain integrated with smart contracts can be used for developing decentralized applications which can address certain real-world problems in a secure and trustworthy manner.
        
    \end{itemize}

	\section{High Level View of The Proposed System}
	\begin{figure*}[ht]
		\begin{center}  
			\includegraphics[scale=0.5]{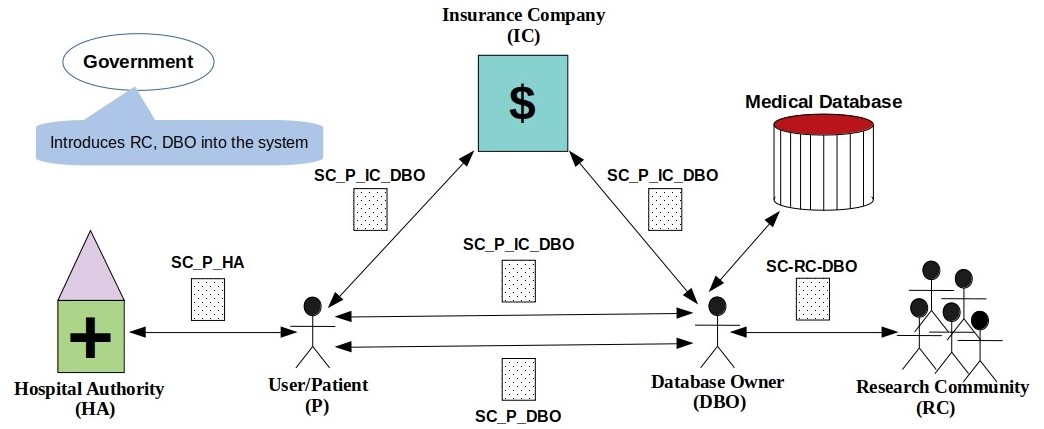}
			\setlength{\belowcaptionskip}{-10pt}
			\caption{\small \sl Healthcare System Model}
				\label{fig:Image2}
		\end{center}  
	\end{figure*}
	\subsection{System Model}
	The major actors or parties involved in our proposed system, as shown in Figure~\ref{fig:Image2}, are as follows -
	
	\textbf{1. Patient (P)/User}
	
	\textbf{2. Hospital Authority (HA)}
	
	\textbf{3. Insurance Company (IC)}
	
	\textbf{4. Database Owner (DBO)}
	
	\textbf{5. Research Community (RC)}
	
	We need to ensure that every party communicates with other parties in a secure and trustworthy manner. There are some fixed protocols or rules that are rigorously defined in the system, and every communicating party should follow this protocol. The protocol suite is written in the form of functions inside the smart contracts
	
	In our system, we have the following smart contracts.
	
	\begin{itemize}[leftmargin=*]
		\item Smart Contract between Patient (P) and Hospital Authority (HA): \textbf{SC\_P\_HA}
		\item Smart Contract between Patient (P), Insurance Company (IC) and Database Owner (DBO): \textbf{SC\_P\_IC\_DBO}
		\item Smart Contract between Patient (P) and Database Owner (DBO): \textbf{SC\_P\_DBO}
		\item Smart Contract between Research Community (RC) and Database Owner (DBO): \textbf{SC\_RC\_DBO}
	\end{itemize}

	We also have one additional smart contract, named \textbf{SC\_Registration}, where the entities of the systems can register themselves before participating in the protocol. Therefore, we have five smart contracts in total to build up the entire system.
	In the next few subsections, we will discuss how these different actors interact among themselves in the system. Before that, let us state some crucial facts or assumptions of our system.
	
	\subsection{Objective}
	We intend to propose a \emph{patient-centric hospital management system} which realizes the following objectives:
	\begin{itemize}
	    \item \emph{Fairness}: An honest party will never lose money even if the rest of the parties are malicious and try to cheat and claim money without providing the desired service or data.
	    \item \emph{Privacy}: Any party cannot view a patient's data until and unless it gets the consent from the patient. 
	    \item \emph{Data Security}: Data of a patient stands protected and cannot be tampered.
	\end{itemize}
	
	\subsection{Assumptions}
    	\begin{enumerate}[label=\alph*)]
    	    \item In our system, every single party has a unique ID. A smart contract generates these IDs at the time of registration in the system. One unique ID corresponds to a particular \emph{PublicKey} To be a part of the system, it is mandatory to register first. There is no option to bypass the \emph{Registration Phase} and get the benefits from the system.
    	    \item In case of the patient claims for medical insurance to reimburse the medical expenditure, he/she must keep the files(related to bills) in the medical repository. Access to these data is strictly restricted. The access control regulation is in the hands of the patient (i.e., data owner). If the patient grants permission to the third party (say, hospital/ Insurance Company, etc.) for accessing his/her records, then only they can do so. The patient can also revoke permission from these entities. The access control matrix is stored in the Blockchain.
    	    \item  In the proposed system, every patient maintains their medical files digitally, i.e., Electronic Health Records (EHR), and these records are stored in a medical database, owned by the Government (or Trusted Third Party Vendor). Thus, the Medical Data Repository Owner or simply the Database Owner (DBO) is considered semi-trusted in our system. A patient can also keep their records to their private storage locally or can appeal to the Database vendor for removing his/her records in case of privacy concerns, which essentially supports the European General Data Protection Regulation (GDPR). As the data is stored in plain text form, DBO can have open access to the stored data. However, this assumption can be removed by adding the scope of handling encrypted data in the database. The model can be changed accordingly in that case. The DBO's activity log is maintained in the blockchain. In case DBO misbehaves, then it can be questioned and penalized accordingly.
    	    \item  In our system, the blockchain is mainly used to store the following -  hash of the records (EHRs) \slash payment bills \slash money receipts, timestamps of asset creation, signatures of issuing authority (as well as the signature of the data owner sometimes) and some other auxiliary pieces of information (context-specific) for supporting accountability, integrity, authentication, and fairness.
    	    \item  Database Owner (DBO) or Research Community (RC) must satisfy certain prerequisite conditions to be a part of the system and appeal to the Government expressing their interest. The conditions or criteria may vary for different Governments of various countries. If all the necessary criteria get satisfied by them, the Government introduces DBO or RC into the system.
    	\end{enumerate}
	
	\subsection{Communication Protocol between Patient and Hospital}
	
	\begin{figure*}
		\begin{center}  
			\includegraphics[scale=0.45]{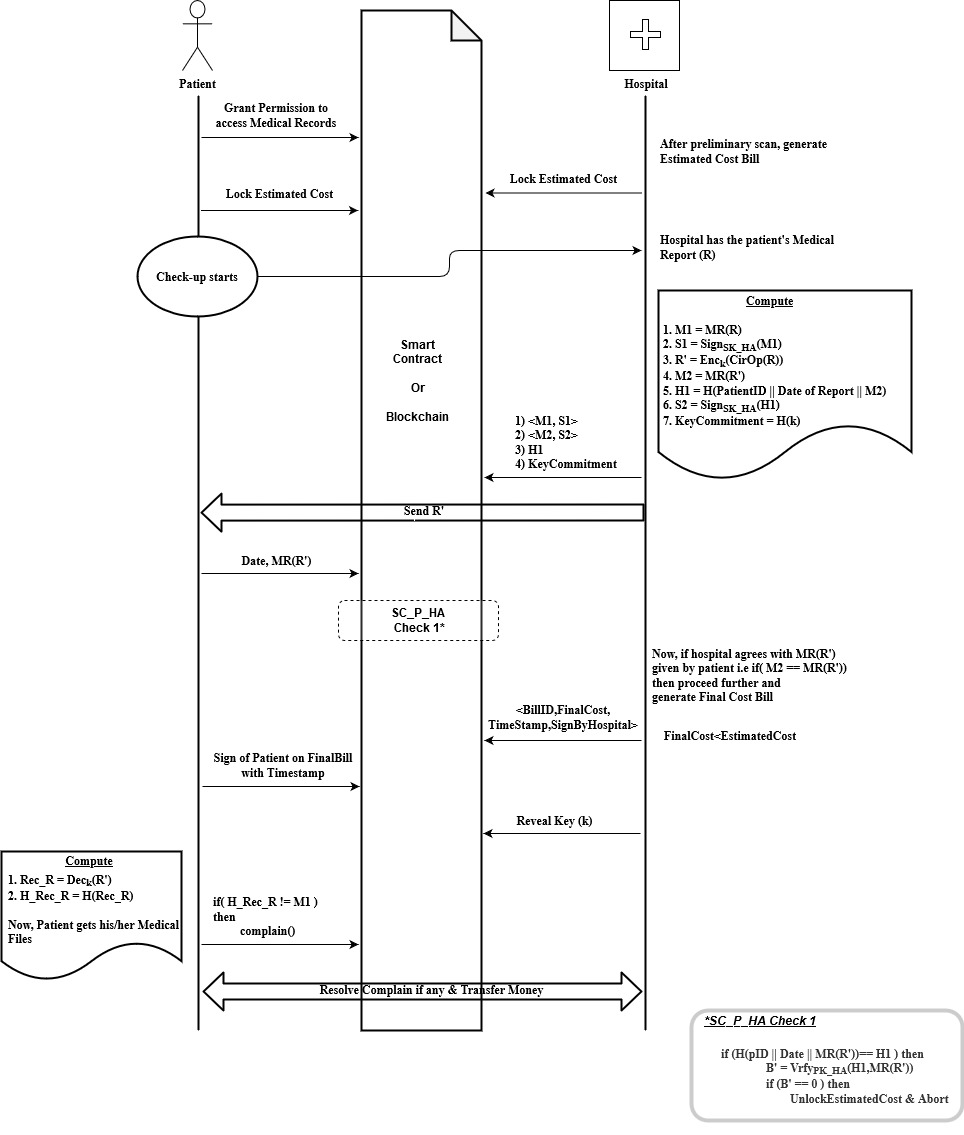}
			\setlength{\belowcaptionskip}{-10pt}
			\caption{\small \sl Interaction between Patient and Hospital
				\label{fig:Image3}}
		\end{center}  
	\end{figure*}
	
	When a patient comes to a hospital with medical issues, the hospital analyzes the patient's problems. After preliminary scanning, the hospital generates an approximate cost of the treatment (i.e., estimated cost). Hospital authority and patient both need to lock this amount to the smart contract (SC\_P\_HA), ensuring fairness in the protocol.
	
	In case the hospital needs to check the patient's medical history from the medical repository; the hospital authority asks the patient to grant proper access permission. The hospital can read the patient's medical history if permission is granted. The patient can revoke access permission from the hospital at any instant of time if needed. Information related to this access control (who has access right to whose data) is stored in the blockchain.
	
	
	Once the estimated cost is locked by the patient, the next obvious task is to start the treatment of the patient within some fixed time window. If the hospital fails to start treatment, then the patient can unlock his or her money after the time window. The hospital invokes a specific function to register the timestamp of starting the treatment.
	
	The patient's testimonials, reports, prescriptions, etc. (i.e., medical files) are generated by the hospital. But these documents are not yet transferred to the patient immediately due to security reasons, which is addressed in \emph{Section 5}. With the help of some cryptographic computations and fair exchange protocol, as shown in Figure~\ref{fig:Image3},
	the hospital sends the medical files to the patient. The hospital stores the following crucial attributes as the metadata in the blockchain.
	
	\begin{enumerate}
		\item Merkle root of the file
		\item Merkle root of the encrypted file
		\item Signature of the hospital on Merkle root of the file.
		\item Signature of the hospital on H(Patient ID || Date of Report || Merkle root of Encrypted File).
	\end{enumerate}
	
	Two Merkle roots are constructed: one with the file chunks as the input, termed as $M_1$ and the other having encrypted chunks of the same file as input, denoted as $M_2$. These Merkle roots are used for verification done by the protocol on behalf of the patient and for verification to be done by other associated entities like Insurance Company and Database Owner. The signature on the Merkle root of the file makes the hospital accountable for the file it has encrypted. The last one in the list, signature on the hash of the patient's attributes and the hash of the encrypted file, adds much more accountability, giving the patient the chance to raise a complaint if the hospital has misbehaved.
	
	Upon receiving these file attributes offline, the patient verifies and gives consent. If the patient finds there is a mismatch in the file attributes, he/she can withdraw the locked amount. If the attributes match, the patient invokes a function to give consent and sign on the file.
	
	In between, the hospital is required to provide the final medical bill to the patient. We assume that the cost of the final bill is not greater than that of the estimated cost bill. The patient has two options - either give consent or raise a dispute for an overcharge of the treatment. With the latter option, the hospital and patient come to a mutual conclusion on the price through offline communication. And this pathway involves two additional transactions to be done by the parties before agreeing to the revised final bill.
	
	Receiving the patient's consent on the final bill, the hospital sends the key to the patient for decrypting the encrypted file that was sent earlier. If the Merkle root of the decrypted file does not match with the one in the contract, the patient raises a complaint by providing a \emph{Proof of Misbehavior}. Once the complaint is verified and the counterparty is found to be malicious, he or she is penalized accordingly. Vice versa, if a wrong complaint has been raised, the party gets penalized.
	
	If the patient neither raises a complaint nor gives consent for the medical report within the timeout period, the hospital withdraws the locked money (patient's and hospital's) and aborts the protocol. In all the functions, there is a timeline check to ensure that each process in the protocol is done within the allocated time window. Also, at each stage of the protocol, both parties are given functions to abort from the protocol to avoid indefinite waiting if one of the parties stops responding. The aforesaid communication model between patient and hospital is depicted in Figure~\ref{fig:Image3}.
	
	\begin{figure*}
		\begin{center}  
			\includegraphics[scale=0.42]{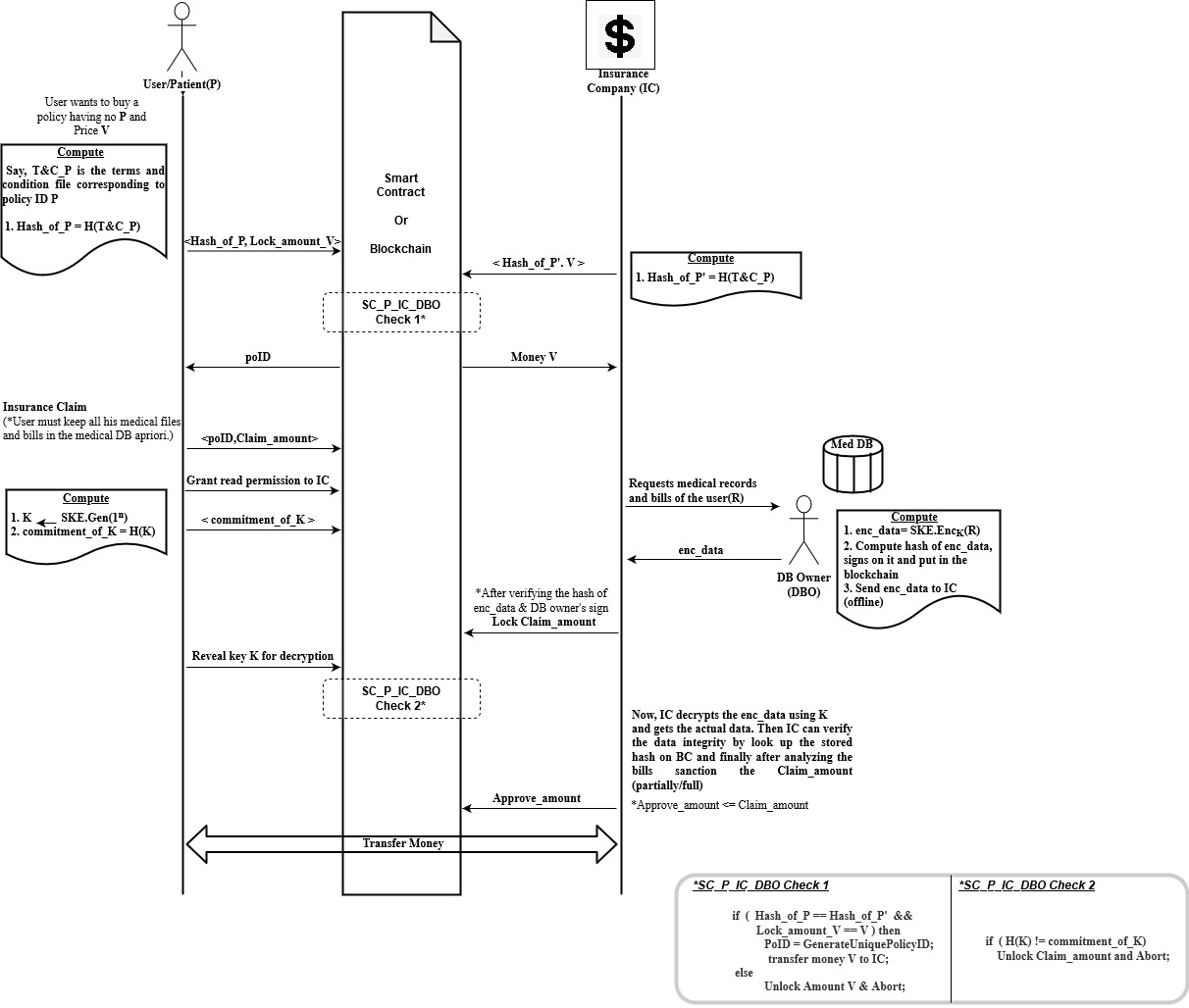}
			\setlength{\belowcaptionskip}{-10pt}
			\caption{\small \sl Interaction among Patient, Insurance Company and Database Owner
				\label{fig:Image4}}
		\end{center}  
	\end{figure*}
	
	\begin{figure*}[ht]
		\begin{center}  
			\includegraphics[scale=0.45]{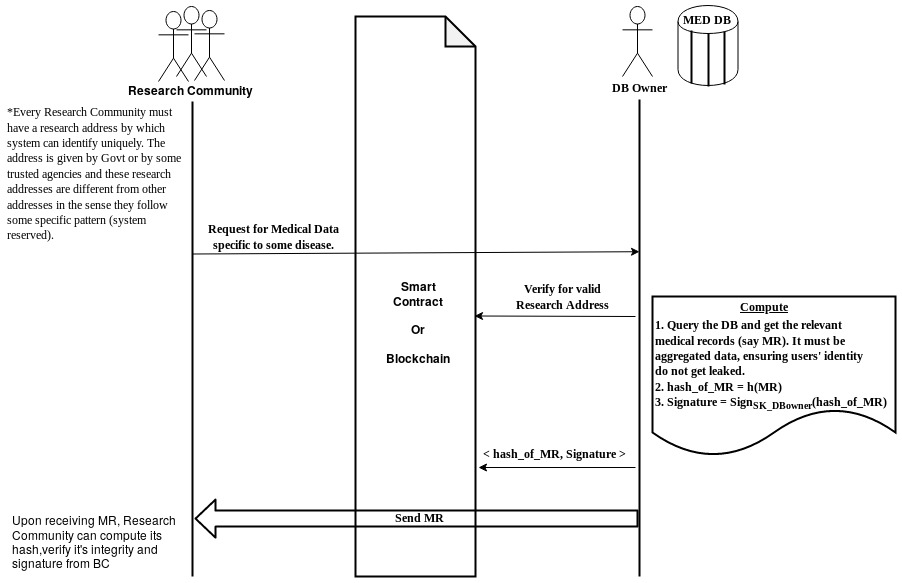}
			\setlength{\belowcaptionskip}{-10pt}
			\caption{\small \sl Interaction among Government, Research Community and Database Owner
				\label{fig:ImageRC}}
		\end{center}  
	\end{figure*}
	
	\subsection{Communication among  Patient, Insurance Company and Database Owner}
	
    We assume that every \emph{Medical Insurance Company} maintains a database locally, where various policy-related information is stored. For simplicity, we can assume that the insurance company maintains a relational table, having schema Policy (policyNo, termsAndcondition, price), policyNo being the primary key. This database is publicly readable.
	
	For policy buying (or claiming) purposes, buyers, or insurance companies would undergo some transactions that are kept in blockchain for future accountability purposes, bringing fairness into the system. Let us denote policyNo, termsAndCondition, and price as $P$, $T\&C\_P$, and $V$ respectively. Buying an insurance policy is a two-phase process. In phase one, the buyer specifies his/her interest to the insurance company for a particular policy ($P$).  The buyer locks the policy price ($V$) in the smart contract and also puts the hash of the terms and condition file ( $H(T\&C\_P)$ ) in the blockchain. In phase two, the insurance company also needs to mention the hash of the terms and conditions file and the policy price to the contract within a fixed time window. If the insurance company responds within this time limit by specifying $H(T\&C\_P)$, and policy price on the blockchain, then smart-contract verifies if these two hash values (i.e, the hash value mentioned by the buyer in phase one and so by the insurance company in phase two) are the same. The contract also checks for the mismatch between the policy price locked by the user and the price mentioned by the insurance company. If they match, then the policy price ($V$) is transferred to the insurance company's account, and the system provides a unique policy ID ($poID$) to the user for future references. All the necessary information corresponding to this $poID$ (e.g., buyer and seller info, the hash of $T\&C\_P$, timestamp of buying, etc.) are stored in the blockchain. From now on, the user/buyer is a legitimate person to get the benefits of the policy upon satisfying the terms and conditions of the policy. In case the two hashes do not match, or the insurance company does not reply within the specific time window, then the user can withdraw his/her locked money from the system. 
	
	The next scenario is Insurance Claiming Procedure. We have proposed a solution with the aid of blockchain, by which an insurance company can verify the claims raised by the user (or patient) for reimbursement of medical bills are correct or not. When the user claims for his/her medical policy, the insurance company asks the user to grant read permission to the information related to the medical expenditure from the Medical Database (assuming the claimant has uploaded all the files related to the medical treatment). The patient's medical information is divided into three categories -
	\begin{enumerate}
	    \item General Information (like name, address, age, etc.)
	    \item Medical Treatment Related Information (like test reports, prescriptions, etc.)
	    \item Medical Expenditure Related Information (like medical bills, invoices, etc.)
	\end{enumerate}
	
	Dividing information into these granular levels helps to provide the information to a third party in a restricted manner. The user grants the necessary permission to the insurance company for reading his/her data from the medical database. Upon getting proper access permission from the user, the Insurance company asks the Database Owner (DBO) to provide the relevant files/records of the user. Here, Database Owner first verifies whether the insurance company has proper permission to access this information. Upon receiving the necessary file(s), the insurance company verifies whether the terms and conditions of the policy are met, and accordingly approves the fund. At this stage, it will be a foolish decision to hand over the data in plaintext format to the Insurance Company. It might just steal the data without reimbursing any money to the patient. To prevent such an attack, we propose a solution. First, the user will generate a one-time secret key $K$ and communicate it with the database owner and keep the commitment of the key in the blockchain. When the Insurance Company asks for the patient's data from DBO, DBO encrypts the data using the key $K$ and sends the encrypted data to the insurance company (offline), and keeps the hash of the encrypted data along with the digital signature on the blockchain. Receiving the encrypted file, the Insurance Company asks the user to provide the key $K$ for decryption. The user only reveals the key if he/she finds that the insurance company has already locked the claimed amount in the smart contract. Having received the key $K$ from the user, the Insurance Company can decrypt the encrypted information. Then the Insurance Company verifies the correctness of the file with the help of blockchain and approves the amount to be sanctioned within some fixed time window. By fetching the stored hash value from the blockchain, the Insurance Company can verify the integrity of the information. The Insurance Company can also check the authenticity of the medical bills by verifying the Digital Signatures (of Hospital Authority and Patient), which are also stored in the blockchain.
	
	An Insurance Company may choose not to respond at all when a user raises a claim request, although the Insurance Company has actively participated in the policy selling process to gain the user's money. This is certainly a malicious behavior of the Insurance Company, which tries to cheat the buyer. We mitigate this sort of risk by allowing the Insurance Companies to lock a certain amount as the security deposit in the system. The security money should not be lesser than the price of the most expensive policy which the company offers to the user (i.e., Security Money >= Costliest Policy Price). An insurance company can deposit or withdraw its security money at any time, however, the company needs to maintain the minimum threshold amount locked all the time. The security deposit compensates the buyer to some extent. If it is not possible to provide the claimed amount due to insufficient security money, the system would try to refund the policy price at least, and even it is not possible to refund the policy price, the insurance company gets deregistered from the system. The activity sequence among patient, insurance company, and database owner is depicted in Figure~\ref{fig:Image4}.
	
	
	\subsection{Communication among Government, Research Community and  Database Owner}
	
	If any Research Community requests access to medical data for research purposes, the system must ensure that the data provided must not be personally identifiable. It means that they will get aggregated data, but they do not have any clue about the patient's identity who is suffering from any health issue. It can be realized in the following way- Research Communities are provided some special addresses (say, addresses starting with some special prefix) by the trusted authority or the Government so that when the research community asks for some data from a database owner, the database owner can easily identify the research group. After this authentication, the database owner provides the aggregated data for specific research interests. The database owner also provides the hash of the aggregated data, digital signature over it, and puts them in the blockchain, so that the research community has some level of confidence about the given medical data. The activity sequence between government, research community, and database owner is depicted in Figure~\ref{fig:ImageRC}.
	
	\section{Implementation and Technical Details}
	In this section, we will depict the algorithms/procedures for the major functions of our smart contracts.
	
	\subsection{Terminology}
	Prior to describing the actual algorithms, let us discuss some important terminologies and notations that we have used in our implementation (Table~\ref{tab:Table1}).
	
	\begin{table}[!ht]
		\centering
		\caption{Terminology \& Notation used in our Scheme}
		\label{tab:Table1}
		\scalebox{0.8}{
			\begin{tabular}{r|l}
				\hline \hline
				\textbf{Abbreviation} & \textbf{Interpretation} \\ \hline \hline 
				BC & Blockchain \\ \hline
				SC & Smart Contract \\ \hline
				P & Patient \\ \hline
				HA & Hospital Authority \\ \hline
				IC & Insurance Company \\ \hline
				DBO & Database Owner \\ \hline
				RC & Research Community \\ \hline
				pAddr & Patient Address \\ \hline
				hAddr & Hospital Address \\ \hline
				icAddr & Insurance Company Address \\ \hline
				dboAddr & Database Owner Address \\ \hline
				rcAddr & Research Community Address \\ \hline
				pID & Patient ID \\ \hline
				hID & Hospital ID\\ \hline
				icID & Insurance Company ID \\ \hline
				dboID & Database Owner ID \\ \hline
				rcID & Research Community ID \\ \hline
				ebID & Estimated Cost Bill ID \\ \hline
				fbID & Final Cost Bill ID \\ \hline
				msID & Multi-Signature ID \\ \hline
				poID & Policy ID \\ \hline
				cID & Claim ID \\ \hline
				rdID & Research Data ID \\ \hline
				asID & Application for Storing ID \\ \hline
				$T_{Event}$ & Timestamp, when the Event occurs \\ \hline
				MR\_File & Merkle Tree Root Hash of File \\ \hline
				CST & Current System Time\\ \hline
				TTL & Threshold Time Limit \\ \hline
			\end{tabular}
		}
	\end{table}
	
	Table ~\ref{tab:Table1} is not exhaustive. Other abbreviations used in the paper are self-explanatory.
	The readers can refer to Table \ref{tab:TableStruct} for the various structure definitions used in the Algorithms (for more details,  the reader may refer to \emph{\textbf{Appendix B}} for various programming constructs - structures, arrays, mappings).
	\begin{table}[!ht]
		\centering
		\caption{Structure Definition}
		\label{tab:TableStruct}
		\scalebox{0.8}{
			 \begin{tabular}{p{0.32\linewidth}|p{0.68\linewidth}}
				\hline \hline
				\textbf{Structure Name} & \textbf{Member Variables} \\ \hline \hline
				EstimatedCheckUpCost & 
				ebID, pID, hID, estimatedCost, $T_{Estimate}$, $T_{LockingByHA}$, $T_{LockingByP}$, $T_{CheckUpStart}$, $T_{UnlockingByHA}, T_{UnlockingByP}$ \\ \hline
				
				FinalCheckUpCost & 
				fbID, pID, hID, finalCost, $T_{FinalBilling}$, $T_{ComplaintByP}$, $T_{UnlockingByHA}$, $T_{UnlockingByP}$, $T_{FinalConsentByP}$\\ \hline
				
				MultiSigOnMedicalData &
				msID, pID, hID, MR\_MedData, MR\_EncCircuitOperatedMedData, H\_x,
				$Sign_{SK_{HA}}MR\_MedData$,
				$Sign_{SK_{HA}}x$, $T_{SigningByHA}$, $T_{VerificationByP}$, $T_{UnlockingByHA}$, $T_{UnlockingByP}$\newline 
		    	*x=(pID||Date||MR\_EncCircuitOperatedMedData)
				\\ \hline
				FileProperties &
				fileSize, fileChunkSize, depth\\ \hline
				KeyAndExchange & 
				key, keyHash, $T_{KeyReveal}$\\ \hline
				ApplicationForStoring & asID, pAddr, dboAddr, msID, key, MR\_File, MR\_EncFile, $T_{Application}$, $T_{VerificationMR}$, $T_{KeyReveal}$, $T_{Complain}$, $T_{Approval}$, $T_{UnlockingByP}$, $T_{UnlockingByDBO}$\\ \hline
				PolicyDetails & 
				poID, buyerID, icID, $T_{BuyingPolicy}$, terms\&ConFileHash, claimIDs[ ]
				\\ \hline
				ClaimDetails &
				cID, ebID, $T_{GeneratingClaimByP}$, claimedAmount,
				approvedAmount, $comm_{K}$, K,
				$T_{RevealKey}$, $T_{LockingByIC}$, $T_{UnlockingByIC}$,
				$T_{Approval}$
				\\ \hline
			\end{tabular}
		}
	\end{table}
	
	\subsection{Algorithms}
	We have classified the algorithms into three categories -
	\begin{enumerate}
	    \item \textbf{Algorithms for Entity Registration}
	    \item \textbf{Algorithm for Data Upload and Access Control}
	    \item \textbf{Algorithm for Data Access and Incentives }
	\end{enumerate}
	
	\smallskip
	\noindent\textbf{A. Algorithms for Entity Registration:} Every entity has a <SK, PK> pair, where PK and SK act as the address and authentication factor of the entity, respectively (Table~\ref{tab:TableKeyGen}).
		
		\begin{table}[!ht]
			\centering
			\caption{Key Generation}
			\label{tab:TableKeyGen}
			\scalebox{0.8}
			{
		    	\begin{tabular}{m{0.08\linewidth}|m{0.42\linewidth}|m{0.50\linewidth}}
		    	     
					\hline \hline
					\textbf{Entity} & \textbf{KeyGen: \newline $(SK,PK) \leftarrow \textrm{ASKE.Gen}(1^n)$} & \textbf{Comments}\\ \hline \hline
					P & ($SK_P , PK_P$) & Key-pair used by Patient.\newline $PK_P$ can also be treated as pAddr. \\ \hline
					HA & ($SK_{HA} , PK_{HA}$) & Key-pair used by Hospital Authority.\newline$PK_{HA}$ can also be treated as hAddr. \\ \hline
					IC & ($SK_{IC} , PK_{IC}$) & Key-pair used by Insurance Company. $PK_{IC}$ can also be treated as icAddr. \\ \hline
					DBO & ($SK_{DBO} , PK_{DBO}$) & Key-pair used by Database Owner.$PK_{DBO}$ can also be treated as dboAddr. \\ \hline
					RC & ($SK_{RC} , PK_{RC}$) & Key-pair used by Research Community. $PK_{RC}$ can also be treated as rcAddr. \\ \hline
				\end{tabular}
			}
		\end{table}
		
		The patient can register his or her name in the Healthcare System by calling the function \textbf{PatientRegistration} (Algorithm~\ref{algo:Algo1}). The parameter $hash\_of\_Personal\_Info \leftarrow \ H(name \vert\vert age \vert\vert mob \vert\vert address)$ is a hash value. It helps to preserve the patient's privacy. Once the registration is done, the patient receives a unique identifier.
		
		
		 \begin{algorithm}[!htb]
                \small
                \DontPrintSemicolon
                \SetKwProg{Fn}{Function}{:}{}
                \SetKw{KwEnd}{end}
                \SetKwFunction{PatientRegistration}{PatientRegistration}
                \Fn{\PatientRegistration{$PK_P$, $hash\_of\_Personal\_Info$}}{
                \If{ ( function invoker address != $pAddr$ OR $PK_P$ is already registered )}{
                    Exit\;
                }
                Generate a unique $pID$\;
                make a mapping or association between $PK_P$ and corresponding $pID$\;
                Record the $hash\_of\_Personal\_Info$ corresponding to $pID$ in $\mathbf{BC}$\;
                \KwEnd
                \;
                }
                \caption{Patient Registration Function}
                \label{algo:Algo1}
            \end{algorithm}

            Similarly, other entities can also register in the system - \\
            \textbf{HA}: $hID \leftarrow \ \textrm{HospitalRegistration}(PK_{HA}, name$)\\
            \textbf{IC}: $icID \leftarrow \ \textrm{InsuranceCompanyRegistration}(PK_{IC}, name$)\\ \textbf{DBO}: $dboID \leftarrow \ \textrm{DatabaseOwnerRegistration}(PK_{DBO}, name$)\\ \textbf{RC}: $rcID \leftarrow \textrm{ResearchCommunityRegistration}(PK_{RC}, name$)\\
            A \textbf{DBO} or an \textbf{RC} cannot register themselves directly into the system. A trusted third party or TTP (here, we consider Government as TTP) is responsible to introduce a DBO or RC into the system. They will call these two functions (i.e., $\textrm{DatabaseOwnerRegistration}$ and $\textrm{ResearchCommunityRegistration}$). 
		    
		\smallskip
		    
		\noindent\textbf{B. Algorithms for Data Upload and Access Control:} Digital medical data or Electronic Health Records (EHR) is generated by the hospital and then transferred to the patient (upon successful discharge and payment of final cost bill). We will discuss later how the hospital generates the EHRs and transfer the same to the patient securely. When the treatment gets completed successfully, either the patient can store the medical files in his/her local storage devices or can decide to store them in some medical repository/database/cloud server. If the patient raises a request for claiming insurance on the medical bill, then the EHRs must be kept in a public medical repository/database, owned by the semi-trusted third-party vendor (also known as DBO or Database Owner). The patient needs to store his/her medical records in the database, and DBO provides the storage space service for some charges. 
		Any two-party fair exchange protocol can be used, however, in this protocol we implement the \textbf{Faiswap Protocol} \cite{dziembowski2018fairswap}, denoted as \textbf{SC\_P\_DBO}, for ensuring a fair exchange of information between patient and DBO. If the DBO finds that the file provided is not correct, it will not provide storage space and abort. The member variables of structure ApplicationForStoring are populated at various invocations of the protocol and these values help in instantiating the two-party exchange protocol. An ID (asID) for applicationForStoring is generated when the patient decides to store the file given by the hospital earlier and invokes \textbf{SC\_P\_DBO} between these two parties. 

	    \smallskip
	    
		\noindent\textbf{C. Algorithms for Data Access and Incentives: }
		\begin{algorithm}[!htb]
                \small
                \DontPrintSemicolon
                \SetKwProg{Fn}{Function}{:}{}
                \SetKw{KwEnd}{end}
                
                \SetKwFunction{GenerateEstimatedCostBill}{GenerateEstimatedCostBill}
                \Fn{\GenerateEstimatedCostBill{$pID$, $estimatedCost$}}{
                \If{ (function caller ID != $hID$) } {
                    Exit
                }
                \If{ (ether value passed with function call != $estimatedCost$) } {
                    Exit
                }
                Generate a unique $ebID$\;
                Instantiates a structure  $EstimatedCheckUpCost\ ec$\;
                Lock the $estimatedCost$ in the $\mathbf{SC}$\;
                Record the CST (say, $T1$) in $\mathbf{BC}$\;
                }
                \KwEnd
                \;
                
                \SetKwFunction{LockEstimatedAmount}{LockEstimatedAmount}
                \Fn{\LockEstimatedAmount{$hID$, $ebID$}} {
                \If{ (function caller ID != $pID$) } {
                    Exit
                }
                Retrieve the structure from $\mathbf{BC}$ : $EstimatedCheckUpCost\ ec$ (corresponding to $ebID$)\;
                \If{ ( ($CST$ - $T1$) > TTL OR (ether value passed with function call != $ec.estimatedCost$) )}{
                     Exit 
                }
                Lock the $estimatedCost$ amount in $\mathbf{SC}$\;
                Record the CST (say $T2$)in $\mathbf{BC}$\
                }
                \KwEnd
                \;
                
                \caption{Pre-Treatment Functions}
                \label{algo:Algo2}
            \end{algorithm}
            
            \begin{algorithm}
                \small
                \DontPrintSemicolon
                \SetKwProg{Fn}{Function}{:}{}
                \SetKw{KwEnd}{end}
                
                \SetKwFunction{StartTreatment}{StartTreatment}
                \Fn{\StartTreatment{$pID$, $ebID$}}{
                \If{ (function caller ID != $hID$) } {
                    Exit
                }
                Generate the medical files\;
                Record the CST (say $T3$) in $\mathbf{BC}$\
                }
                \KwEnd
                \;
                
                \SetKwFunction{KeepSignedHashToBlockchain}{KeepSignedHashToBlockchain}
                \Fn{\KeepSignedHashToBlockchain{$pID$, $MR(MedData)$, $MR(Enc(CircuitOperatedMedData))$, $H(pID||Date||Enc(CircuitOperatedMedData))$,\newline
            $Sign_{SK_{HA}}(H(pID||Date||Enc(CircuitOperatedMedData)))$, $Sign_{SK_{HA}}(MR(MedData))$, $H(key)$}} {
                \If{ (function caller ID != $hID$) } {
                    Exit
                }
                Keep $MR(MedData)$ and $MR(Enc(CircuitOperatedMedData))$ in $\mathbf{BC}$\;
                Keep $Sign_{SK_{HA}}(MR(MedData))$ and $Sign_{SK_{HA}}(H(pID||Date||Enc(CircuitOperatedMedData)))$ in $\mathbf{BC}$\;
                Store $H(key)$ in $\mathbf{SC}$\;
                Record the CST (say, $T4$) in $\mathbf{BC}$\;
                }
                \KwEnd
                \;
                
                \SetKwFunction{VerifyAndGiveConsent}{VerifyAndGiveConsent}
                \Fn{\VerifyAndGiveConsent{$msID$, $H(pID||Date||Enc(CircuitOperatedMedData))$}}{
                \If{ (function caller ID != $pID$) } {
                    Exit
                }
                \If{ ($CST$ - $T4$) > TTL)}{
                     Exit 
                }
                Verify the two hashes and if their respective signatures were signed by the right HA\;
                Record the CST (say, $T5$) in $\mathbf{BC}$\;
                }
                \KwEnd
                \;
                
                \SetKwFunction{DischargeAndGenerateFinalCostBill}{DischargeAndGenerateFinalCostBill}
                \Fn{\DischargeAndGenerateFinalCostBill{$ebID$, $pID$, $finalCost$}}{
                \If{ (function caller ID != $hID$) } {
                    Exit
                }
                \If{ ($CST$ - $T5$) > TTL)}{
                     Exit 
                }
                Get $estimatedCost$ corresponding to $ebID$\;
                \If{ ($finalCost$>$estimatedCost$) } {
                    Exit
                }
                Instantiate the $FinalCheckUpCost$ struct and update the price\;
                Record the CST (say, $T6$) in $\mathbf{BC}$\;
                }
                \KwEnd
                \;
                
                \SetKwFunction{ConsentFinalBillPatient}{ConsentFinalBillPatient}
                \Fn{\ConsentFinalBillPatient{$fbID$, $hID$}}{
                \If{ (function caller ID != $pID$) } {
                    Exit
                }
                \If{ ($CST$ - $T6$) > TTL)}{
                     Exit 
                }
                Record the CST (say, $T7$) in $\mathbf{BC}$\;
                }
                \KwEnd
                \;
                
                \caption{Functions for Generating Medical File and Final Bill}
                \label{algo:Algo3}
            \end{algorithm}
                
            \begin{algorithm}
                \small
                \DontPrintSemicolon
                \SetKwProg{Fn}{Function}{:}{}
                \SetKw{KwEnd}{end}
                
                \SetKwFunction{KeyReveal}{KeyReveal}
                \Fn{\KeyReveal{$pID$, $ebID$, $key$}}{
                \If{ (function caller ID != $hID$) } {
                    Exit
                }
                \If{ ($CST$ - $T7$) > TTL)}{
                     Exit 
                }
                \If{ ($Hash(key)$ in $\mathbf{BC}$ != $Hash(key)$) } {
                    Exit
                }
                Store the key in the $\mathbf{BC}$\;
                Record the CST (say, $T8$) in $\mathbf{BC}$\;
                }
                \KwEnd
                \;
                
                \SetKwFunction{PatientFinalConsent}{PatientFinalConsent}
                \Fn{\PatientFinalConsent{$ebID$, $hID$}} {
                \If{ (function caller ID != $pID$) } {
                    Exit
                }
                \If{ ($CST$ - $T8$) > TTL)}{
                     Exit 
                }
                Record the CST (say, $T9$) in $\mathbf{BC}$ as $T_{FinalConsentByP}$\;
                }
                \KwEnd
                \;
                
                \SetKwFunction{PatientComplain}{PatientComplain}
                \Fn{\PatientComplain{$complaint$, $encodedVectorsThisGate$, $gateIndexes$, $ebID$, $hID$}}{
                
                Get $fbID$, $msID$ from the $ebID$\;
                Get $finalCheckUpCost$, $multiSigOnMedicalData$, $MultiSignIdToEncFileProperties$ structs from the corresponding IDs\;
                
                \If{ (function caller ID != $pID$) } {
                    Exit
                }
                \If{ ($CST$ - $T9$) > TTL)}{
                     Exit 
                }
                \If{ ($T_{FinalConsentByP}$!=0)}{
                     Exit 
                }
                \If{ (length of complaint !=0 length of encodedVectorsThisGate)}{
                     Transfer the final amount to the hospital\;
                     Exit 
                }
                \If{ (length of gateIndexes < maxLinesToGate+1)}{
                     Transfer the final amount to the hospital\;
                     Exit 
                }
                \If{ (MerkleProof(encodedVectors,encFileRootHash)}{
                     Transfer the final amount to the hospital\;
                     Exit 
                }
                Decrypt the encoded input vectors and the encoded output vector of the gate\;
                Calculate the resultant vector from the input vectors and the gate operation\;
                \If{ (Resultant gate output vector != Decrypted gate output vector )}{
                     Transfer the final amount to the hospital\;
                     Exit 
                }
                
                Record the CST (say, $T10$) in $\mathbf{BC}$ as $T_{ComplaintByP}$\;

                }
                \KwEnd
                \;
                
                \caption{Functions for File Consent}
                \label{algo:Algo4}
            \end{algorithm}

        \begin{enumerate}[leftmargin=0cm,itemindent=.5cm,labelwidth=\itemindent,labelsep=0cm,align=left]
            \itemsep2.5mm
            \item \textbf{Generating Medical Reports of Patient by Hospital}: Here, we will discuss the algorithms related to generating the new medical reports of a patient, fairly communicate the same to the patient party preserving the integrity of the files, generating the final cost bill, and finally, discharging the patient. We break it into two phases - \textbf{Pre-Treatment} and \textbf{Generation of Medical File and Final Bill}, described in Algorithm~\ref{algo:Algo2} and Algorithm \ref{algo:Algo3} respectively. 
            \begin{enumerate}[leftmargin=0cm,itemindent=.5cm,labelwidth=\itemindent,labelsep=0cm,align=left]
                \itemsep2mm
                \item \emph{The Algorithms for \textbf{Pre-Treatment} (Algorithm~\ref{algo:Algo2})}\\
                \begin{itemize}
                    \itemsep1mm
                    \item The patient, having ID $pID$, reports to the hospital the medical problem he or she is facing. Hospital, having ID $hID$, estimates the cost for the medical service (price $V$) and locks the amount with the invocation of the \emph{\textbf{GenerateEstimatedCostBill}} function (say, at time $T1$). The function generates a unique estimated cost bill ID $ebID$ and also instantiates $EstimatedCheckUpCost$ with this $ebID$. The $ebID$ is communicated with the patient offline.
                    \item Upon receiving the $ebID$, the patient needs to invoke the function \emph{\textbf{LockEstimatedAmount}} (say, at time $T2$) within a limited time period ($TTL$) to lock the $estimatedCost$ amount.
                \end{itemize}
                
                \item \emph{The Algorithms for \textbf{Generating Medical file and Final Bill } (Algorithm~\ref{algo:Algo3})}\\  
                \begin{itemize}[leftmargin=*]
                    \itemsep1mm
                    \item Upon patient locking the estimated amount, the hospital needs to invoke the function \emph{\textbf{StartTreatment}} (say, at time $T3$) within a limited time period ($TTL$) to start the treatment process. The medical files are generated after the function invocation. Keeping in mind the variable nature of the medical processes, there is no time limit set for the next immediate function (\emph{\textbf{KeepSignedHashToBlockchain}})'s invocation.
                    \item The medical file(MedData) generated is broken into chunks of predetermined size. A Merkle tree is constructed with the chunks and the Merkle root $MR(MedData)$ is obtained. The chunks are operated according to a circuit. In our case, the Boolean circuit maps to the construction of the Merkle tree. Hash is the primary operation of the gates used. In any gate of this circuit, the child gate’s output vectors are concatenated and used as inputs to the gate. The concatenated vector is hashed, and the resultant value is the output of the gate. The gate outputs are encrypted with a key, and all the encrypted chunks are used to construct a different Merkle tree and obtain a Merkle root $MR(Enc(CircuitOperatedMedData))$. This Merkle root is concatenated with the patient ID, date of report, etc. It is hashed and signed $Sign_{SK_{HA}}(H(pID||Date||Enc(CircuitOperatedMedData)))$. Hospital also signs on the Merkle root of the file $Sign_{SK_{HA}}(MR(MedData))$. The two hashes, the corresponding signatures, and few file-related variables are the parameters to the function \emph{\textbf{KeepSignedHashToBlockchain}}. Let us say the function is invoked at the time $T4$.
                    \item The patient by this time might have received the encrypted file chunks from the hospital. The patient constructs the Merkle tree with the received $Enc(CircuitOperatedMedData)$ and checks whether the constructed Merkle root and the root that is kept in the blockchain. Patient also verifies the matching of the concatenated hash $(H(pID||Date||Enc(CircuitOperatedMedData)))$ discussed in the last function. The two signatures - $Sign_{SK_{HA}}(MR(MedData))$ and  $Sign_{SK_{HA}}(H(pID||Date||Enc(CircuitOperatedMed$\newline$Data)))$ are also verified by the patient. If these parameters are valid, then the patient invokes the function \emph{\textbf{VerifyAndGiveConsent}} (say, at time $T5$) within a limited time period ($TTL$).
                    \item If the estimated cost and the actual cost are the same, the hospital invokes the function \emph{\textbf{DischargeAndGenerateFinalCostBill}} (say, at time $T6$) within a limited time period ($TTL$) by keeping the final cost, $finalCost$, same as the estimated cost $estimatedCost$. Otherwise, the hospital puts up a new final cost by invoking the function. Here, the final cost is always less than or equal to the estimated cost because the estimated cost is an overestimation of the treatment cost.
                    \item The patient calls the function \emph{\textbf{consentFinalBillPatient}} (say, at time $T7$) within a limited time period ($TTL$) if the final cost provided by the hospital is satisfactory.
                \end{itemize}

            \end{enumerate}
            
            The Algorithm for \textbf{File Consent} is described in Algorithm~\ref{algo:Algo4}. In this step, the hospital provides the key for decrypting the file. Any misbehavior detected leads to penalization of the party involved in providing wrong or no information.
            \begin{enumerate}[leftmargin=0cm,itemindent=.5cm,labelwidth=\itemindent,labelsep=0cm,align=left]
                \itemsep2mm
            
            \item \emph{The Algorithms for \textbf{File Consent} (Algorithm~\ref{algo:Algo4})}\\  
                \begin{itemize}[leftmargin=*]
                    \item The hospital reveals the key by invoking the function \emph{\textbf{KeyReveal}}  (say, at time $T8$) within a limited time period ($TTL$). Then the key is hashed and checked with the hash of the key $H(key)$ (key's commitment) kept in the contract.
                    \item When the hospital discloses the key in the blockchain, the patient decrypts the file chunks and performs the circuit operation of the gates one by one. If all the checks are valid, then the patient gives consent by invoking the function \emph{\textbf{PatientFinalConsent}} (say, at time $T9$) within a limited time period ($TTL$) and the locked amount is transferred to the hospital.
                    \item If the above-mentioned check produces any mismatch, an appropriate complaint is generated, and it is produced by the patient through the function \emph{\textbf{PatientComplain}} (say, at time $T10$) within a limited time period ($TTL$). If the complaint is a valid one, the hospital is penalized and vice-versa.
                \end{itemize}
            \end{enumerate}
            
            \item \textbf{Accessing client's medical report by Insurance Company in order to approve policy claim.} 
            We can further subcategorize these algorithms into four parts - Procedure for \textbf{Buying an Insurance Policy} fairly from an IC is stated in Algorithm~\ref{algo:Algo5}. Procedure for \textbf{Claiming an Insurance} by a Policy Buyer is stated in Algorithm~\ref{algo:Algo6} and for \textbf{Processing a Claim} is stated in Algorithm~\ref{algo:Algo7}. \textbf{Approving a Claim} has been described in Algorithm~\ref{algo:Algo8}.
                \begin{algorithm}[!htb]
                \small
                \DontPrintSemicolon
                \SetKwProg{Fn}{Function}{:}{}
                \SetKw{KwEnd}{end}
                
                \SetKwFunction{BuyPolicyPhaseOne}{BuyPolicyPhaseOne}
                \Fn{\BuyPolicyPhaseOne{$icID$, $hashOfTermsAndCon$}}{
                Store the $hashOfTermsAndCon$, $icID$ in $\mathbf{BC}$\;
                Lock the Policy Price $V$ in $\mathbf{SC}$\;
                Record the CST (say, $t1$) in $\mathbf{BC}$\;
                }
                \KwEnd
                \;
                
                \SetKwFunction{BuyPolicyPhaseTwo}{BuyPolicyPhaseTwo}
                \Fn{\BuyPolicyPhaseTwo{$pID$, $price$, $hashOfTermsAndCon$}} {
                \If{ (function caller ID != $icID$) } {
                    Exit
                }
                \If{ ($price$ > $securityMoney_{icID}$) } {
                    De-register the IC from the System and Exit\;
                }
                Record the CST (say, $t2$)\;
                \If{ ( ($t2 -t1$) > TTL) } {
                    Exit\;
                }
                Retrieve the amount of money locked by buyer $pID$ from $\mathbf{BC}$ and matches with $price$\;
                Retrieve the hash of terms and condition file mentioned by buyer $pID$ from $\mathbf{BC}$ and matches with $hashOfTermsAndCon$\;
                \If{ (the above matches are found incorrect) }{
                    Exit\;
                }
                Generate a unique $poID$\;
                Instantiate a structure $PolicyDetails$ $pd$ corresponding to the $poID$ and store in $\mathbf{BC}$\;
                \KwRet {($poID$)}\;
                }
                \KwEnd
                \;
                
                \SetKwFunction{WithdrawLockedPolicyBuyingMoney}{WithdrawLockedPolicyBuyingMoney}
                \Fn{\WithdrawLockedPolicyBuyingMoney{$icID$}}{
                Retrieve the timestamp($t1$) of locking the money from $\mathbf{BC}$\;
                \If{ ($CST$ - $t1$) > TTL) AND the seller IC does not respond }{
                     Unlock the money from $\mathbf{SC}$ and Credit to the Buyer's Account\; 
                }
                }
                \KwEnd
                \;
                
                \caption{Functions for Buying an Insurance Policy fairly}
                \label{algo:Algo5}
            \end{algorithm}

            \begin{algorithm}[!htb]
            \small
            \DontPrintSemicolon
            \SetKwProg{Fn}{Function}{:}{}
            \SetKw{KwEnd}{end}
            
            \SetKwFunction{ClaimMoney}{ClaimMoney}
            \Fn{\ClaimMoney{$poID$, $ebID$, $asID$, $claimedAmount$, $comm\_K$}}{
            Obtain $fbID$ and $msID$ from $ebID$\;
            Retrieve the structures from $\mathbf{BC}$ : $PolicyDetails$ $pd$, $FinalCheckUpCost$ $fc$ and $ApplicationForStoring$ $as$ (corresponding to $poID$, $fbID$ and $asID$ respectively)\;
            Parse the structures' data and Verify the necessary field values for correctness\;
    	    Check( function caller ID == $pd.buyerID$ == $fc.pID$ == $as.pID$ )\;
    	    Check( $as.multiSigID$ == $msID$ AND $as.timestamp\_approval$ != 0 )\;
    	    Check( $claimedAmount \leq fc.finalCost$ )\;
    	    Embed the business logic to check that the user has not claimed for the same bill before.\; 
    	    \If{ ( any of the above checks is FALSE ) } {
                Exit\;
            }
            Generate a unique $cID$\;
    		Instantiate a structure $ClaimDetails$ $cd$\;
    		$cd.comm_{K} \leftarrow comm\_K$\;
    		$cd.T_{GeneratingClaimByP} \leftarrow CST$\;
    		Store $cd$ into $\mathbf{BC}$\;
    		Make an entry in mapping: $cIDTopoID[cID] \leftarrow poID$\;
    		$GrantReadAccessPermission(pd.icID)$\;
    		}
            \KwEnd
            \;
            \caption{Function for Claiming Insurance }
            \label{algo:Algo6}
            \end{algorithm}

            \begin{algorithm}[!htb]
            \small
            \DontPrintSemicolon
            \SetKwProg{Fn}{Function}{:}{}
            \SetKw{KwEnd}{end}
            
            \SetKwFunction{KeepSigOnHashOfEncFile}{KeepSigOnHashOfEncFile}
            \Fn{\KeepSigOnHashOfEncFile{$cID$, $sign\_DBO$}} {
            Store <$sign\_DBO$ and $CST$> corresponding to $cID$ in the $\mathbf{BC}$\;
    		Mark the key $K$ as expired\;
            }
            \KwEnd
            \;
            
           \SetKwFunction{LockClaimedMoney}{LockClaimedMoney}
            \Fn{\LockClaimedMoney{$cID$, $amount$}}{
            Obtain $poID \leftarrow  cIDTopoID[cID]$\;
    	    Retrieve the structures from $\mathbf{BC}$: $PolicyDetails \ pd$, $ClaimDetails \ cd$ (corresponding to $poID$ and $cID$ respectively)\;
    	    \If{ ( function invoker ID != $pd.icID$ ) }{
    	        Exit\;
    	    }
    	    \If{ ( $amount$ == $cd.claimedAmount$ ) }{
    	        Lock the $amount$ in the $\mathbf{SC}$\;
    		    $cd.T_{LockingByIC} \leftarrow  CST$\;
    	    }
            }
            \KwEnd
            \;
            
            \SetKwFunction{RevealSecretKey}{RevealSecretKey}
            \Fn{\RevealSecretKey{$cID$, $K$}}{
            Obtain $poID \leftarrow cIDTopoID[cID]$\;
    	    Retrieve the structures from $\mathbf{BC}$ : $PolicyDetails$ $pd$, $ClaimDetails$ $cd$ (corresponding to $poID$ and $cID$ respectively)\;
    	    \If{ ( function invoker ID != $pd.buyerID$ ) }{ 
    	        Exit\;
    	    }
    	    \If{ ( H($K$)!= $cd.comm_{K}$ OR\newline ($CST$ - $cd.T_{LockingByIC}$) > TTL )}{ 
    	        Exit\;
    	    }
    	    $cd.K \leftarrow K$\;
    	    $cd.T_{RevealKey} \leftarrow CST$\;
            }
            \KwEnd
            \;
            
            \SetKwFunction{WithdrawLockedClaimedMoney}{WithdrawLockedClaimedMoney}
            \Fn{\WithdrawLockedClaimedMoney{$cID$}}{
            Obtain $poID \leftarrow cIDTopoID[cID]$\;
    	    Retrieve the structures from $\mathbf{BC}$ : $PolicyDetails$ $pd$, $ClaimDetails$ $cd$ (corresponding to $poID$ and $cID$ respectively)\;
    	    \If{ ( function invoker ID != $pd.icID$ ) }{
    	        Exit\;
    	    }
    	    \If{ ( $cd.T_{RevealKey}$ == 0 AND $cd.T_{LockingByIC}$ != 0 AND ($CST$ - $cd.T_{LockingByIC}$) > TTL ) }{
    		    Unlock the $claimedAmount$ and Credit to the IC's Account.
    	    }
            }
           \KwEnd
            \;
           
            \caption{Functions for Processing an Insurance  Claim}
            \label{algo:Algo7}
            \end{algorithm}
            
            \begin{algorithm}[!htb]
            \small
            \DontPrintSemicolon
            \SetKwProg{Fn}{Function}{:}{}
            \SetKw{KwEnd}{end}
            
            \SetKwFunction{ApproveClaim}{ApproveClaim}
            \Fn{\ApproveClaim{$cID$, $approvedAmount$}}{
            Obtain $poID \leftarrow cIDTopoID[cID]$\;
        	Retrieve the structures from $\mathbf{BC}$ : $PolicyDetails$ $pd$, $ClaimDetails$ $cd$ (corresponding to $poID$ and $cID$ respectively)\;
        	\If{ ( function invoker ID != $pd.icID$ ) }{
        	    Exit\;
        	}
        	\If{ ( $cd.T_{RevealKey}$ == 0 OR $cd.T_{Approval}$ != 0 OR \newline ($CST$ - $cd.T_{RevealKey}$) > TTL ) }{
        	    Exit\;
        	}
        	$cd.approvedAmount \leftarrow approvedAmount$\;
        	$cd.T_{Approval} \leftarrow CST$\;
        	Unlock the $approvedAmount$ and Credit to the Buyer's Account\;
        	$remainingAmount \leftarrow (cd.claimedAmount - cd.approvedAmount)$\;
        	\If{ ( $remainingAmount$ > 0 ) }{
        	    Unlock the $remainingAmount$ and Credit to the IC's Account\;
        	}
            }
            \KwEnd
            \;
            
            \SetKwFunction{SelfApproveClaim}{SelfApproveClaim}
            \Fn{\SelfApproveClaim{$cID$, $icID$}}{
            Obtain $poID \leftarrow cIDTopoID[cID]$\;
        	Retrieve the structures from $\mathbf{BC}$ : $PolicyDetails$ $pd$, $ClaimDetails$ $cd$ (corresponding to $poID$ and $cID$ respectively)\;
        	\If{ ( function invoker ID != $pd.buyerID$ ) }{
        	    Exit\;
        	}
        	\If{ ( $cd.T_{RevealKey}$ != 0 AND $cd.T_{Approval}$ == 0 AND ($CST$ - $cd.T_{RevealKey}$) > TTL) }{
        	    $cd.approvedAmount \leftarrow cd.claimedAmount$\;
        		$cd.T_{Approval} \leftarrow CST$\;
        		$cd.isSelfApproved \leftarrow TRUE$\;
        		Unlock the $claimedAmount$ and Credit to the Buyer's Account.
        	}
            }
           \KwEnd
            
            \caption{Functions for Approving an Insurance  Claim}
            \label{algo:Algo8}
            \end{algorithm}
        
            \begin{enumerate}[leftmargin=0cm,itemindent=.5cm,labelwidth=\itemindent,labelsep=0cm,align=left]
                \itemsep2mm
                \item \emph{The Algorithms for \textbf{Buying an Insurance Policy} fairly from an IC (Algorithm~\ref{algo:Algo5})}\\
                \begin{itemize}[leftmargin=*]
                    \itemsep1mm
                    \item Say, a user having ID $pID$ wants to buy a policy from an IC (having ID $icID$). The price of the policy is $V$ unit. The user reads the terms and conditions file ($T\&C\_P$ in fig~\ref{fig:Image4}), where all the policy-related details and benefits are mentioned. Then the user computes the hash of the terms and condition file locally as - \\
                    \(hashOfTermsAndCon \leftarrow H(T\&C\_P)\) \\
                    and invokes the function \emph{\textbf{BuyPolicyPhaseOne}} (say, at time $t1$).  
                    
                    \item In the first phase of buying, the user locks the policy price $V$ in the SC and also keeps the hash of the terms \& conditions file in the BC. In the second phase, the IC should acknowledge the buyer's interest to buy a policy within a fixed time window ($TTL$). So, the IC invokes the function \emph{\textbf{BuyPolicyPhaseTwo}} (say, at time $t2$). IC needs to specify the hash of the terms \& conditions file and the policy price. The smart contract verifies whether these values specified by both parties (specified by policy buyer in phase one and so by the insurance company in phase two) match or not.
                
                    \item After the first phase of buying, if the IC does not respond within the fixed time period (i.e., $(t2-t1)>TTL$ ), the policy buyer can withdraw his or her locked money (policy price) from the system by calling the function \emph{\textbf{WithdrawLockedPolicyBuyingMoney}}.
                \end{itemize}
                
                \item \emph{The Algorithms for \textbf{Claiming an Insurance} by a Policy Buyer(Algorithm~\ref{algo:Algo6})}
                \begin{itemize}[leftmargin=*]
                    \itemsep1mm
                    \item The policy buyer or the user can claim to the IC for reimbursement of his or her medical expenditure by calling the function \emph{\textbf{ClaimMoney}}. The user generates a temporary key $K$  and computes the commitment of the key $K$ offline.\newline 
                    $K \leftarrow SKE.Gen(1^n)$.\newline $comm\_K \leftarrow H(K)$.
                    \newline The user sends the key $K$ to the DBO (offline, one-to-one communication) and specifies the $comm\_K$ (as function parameters) while invoking the function. So, $comm\_K$ would go to the $BC$, once the transaction gets mined. This function essentially authenticates the identity of the user, validates the important data (mentioned as the member variables of various structures), verifies the signatures, and also checks for the timestamps of various key activities with the aid of $BC$. The algorithm also checks that the policy-holder can not claim for the same medical bill more than once. If everything is found right, then the function generates a unique $cID$ and communicates the same to the user for future references. The function also instantiates a structure of type $ClaimDetails$ with necessary values and stores in the $BC$. The User also needs to grant read access permission to the IC for verifying the medical files/bills from the repository.
            
                \end{itemize}
                
                \item \emph{The Algorithms for \textbf{Processing a Claim} (Algorithm~\ref{algo:Algo7})}\\  
            
                \begin{itemize}[leftmargin=*]
                    \itemsep1mm
                    \item Obtaining the Key $K$ from the user, the DBO verifies whether $comm\_K$ matches with the received key. If it matches, then DBO encrypts the user's file (say, the file is R).\newline
                    $enc\_data \leftarrow SKE.Enc_{K}(R)$\newline
                    Once the encryption is done, the DBO must invalidate the key $K$. Invalidation of $K$ is essential, else it leads to security risk. The DBO computes the hash of the encrypted data, signs over it, and puts it into the BC for accountability purpose (i.e., $Sign_{SK_{DBO}}( H(enc\_{data}) )$ ). 
                    The DBO invokes the function \emph{\textbf{KeepSigOnHashOfEncFile}} and sends the encrypted file in offline mode to the IC. 
                    
                    \item Next, the IC needs to lock the claimed $amount$ in the $\mathbf{SC}$ and invokes the function \emph{\textbf{LockClaimedMoney}}.
                
                    \item Once IC locks the claimed amount, the user (i.e., policy buyer) should reveal the key $K$ within some fixed time window($TTL$) so that IC can decrypt the encrypted files. The user calls the function \emph{\textbf{RevealKey}}.
                
                    \item In case the policy claimer (i.e., user) has not provided the key $K$ within the specified time window, the IC can withdraw the locked money by invoking the function \emph{\textbf{WithdrawLockedClaimedMoney}}.
                
                \end{itemize}
                
                \item \emph{The Algorithms for \textbf{Approving a Claim} (Algorithm~\ref{algo:Algo8})}\\  
            
                \begin{itemize}[leftmargin=*]
                    \itemsep1mm
                    \item Getting the key $K$ from the user, the IC decrypts the encrypted file and verifies all the necessary information. After verification, the IC approves the claim as per the norms specified in the \emph{Terms and Conditions} of the policy (i.e., $T\&C\_P$). The claim is not necessarily approved all the time fully, it may also be approved partially. Here, the key point is that - the IC is confined to obey all the terms defined in the $T\&C\_P$ file. There is no way to bypass any rules or norms, as the hash of this file was already stored in the BC. The IC should invoke the function \emph{\textbf{ApproveClaim}} within a fixed time window($TTL$) after the key is revealed by the user.
                
                    \item If the IC does not invoke the above function within the specified time window (even if the user has revealed the key in proper time), we give the user the provision to validate the claim amount entirely by himself (or herself). In such cases, the policy-holder can take away the locked money of the IC by calling the function \emph{\textbf{SelfApproveClaim}}. It takes care of the user's fairness.
            
                \end{itemize}
                    
            \end{enumerate}

        \begin{algorithm}[!htb]
            \small
            \DontPrintSemicolon
            \SetKwProg{Fn}{Function}{:}{}
            \SetKw{KwEnd}{end}
            
            \SetKwFunction{RequestDataForResearch}{RequestDataForResearch}
            \Fn{\RequestDataForResearch{$dboID$, $hash\_Query$}}{
            \If{(function caller ID is not a $rcID$)}{
		        Exit\;
	        }
	        Generate a unique rdID\;
	        Store <$dboID$, $rcID$, $hash\_Query$> to the $\mathbf{BC}$ correspoding to $rdID$\;
	        Record the CST(say,$t_1$) in $\mathbf{BC}$\;
	        
    		}
            \KwEnd
            \;
            
            \SetKwFunction{ProvideDataForResearch}{ProvideDataForResearch}
            \Fn{\ProvideDataForResearch{$rdID$, $hash\_Data$, $sign$}}{
            \If{(function caller ID != $dboID$)}{
		        Exit\;
	        }
	        Fetch the timestamp($t_1$)\ related\ to\ $rdID$\ from\ $\mathbf{BC}$\;
	        
	        \If{ ($CST$ - $t_1$ > TTL) }{
		        Exit\;
	        }
	        
	        Store <$hash\_Data$, $sign$> to the $\mathbf{BC}$ correspoding to $rdID$\;
	        Record the CST(say,$t_2$) in $\mathbf{BC}$\;
	        
    		}
            \KwEnd
            \;
            
            \caption{Functions for Accessing Medical Data by RC}
            \label{algo:Algo9}
        \end{algorithm}
        
        \item \textbf{Providing medical data to RC for research purpose}: When Research Community (RC) seeks medical data, it can directly approach the DBO to provide medical data for the sake of research. We have already depicted the model in section 3.6. Procedure for \emph{Accessing Medical Data by RC} is stated in Algorithm~\ref{algo:Algo9}.
        
            \begin{itemize}[leftmargin=*]
                \itemsep1mm
                \item First, RC must obtain a license from the Government to carry on the Research Activity. The Government has the only authority to get the RC registered in the system. Therefore, the former introduces the RC in the system by calling the function \emph{\textbf{ ResearchCommunityRegistration}}. As a result, the RC will receive a unique rcID.
                \item After getting registered in the system, the RC can ask the DBO to provide medical data for the sake of research by calling the function \emph{\textbf{RequestDataForResearch}}. 
                \item In response to the RC's data request, the DBO provides relevant medical data to the RC. Here, the key point to be remembered is that DBO cannot send any information which can compromise the patient's privacy. DBO can only provide aggregated data but, at the same time, abide by the rules stated to preserve privacy (semi-trusted assumption). The DBO computes the hash of the medical data which is to be sent and also puts a signature on this. These are kept in the BC. The DBO invokes this function \emph{\textbf{ProvideDataForResearch}}.
            \end{itemize}
       
        \end{enumerate}
        Reader may refer to \emph{\textbf{Appendix A}}, where the function (or activity) sequence diagrams are given for the contracts SC\_P\_HA, SC\_P\_IC\_DBO and SC\_RC\_DBO.
        
	\section{Security Analysis}
	Blockchain technology uses some cryptographic primitives (e.g., hash function, digital signature). As long as the underlying cryptographic primitives are secured, the blockchain is secure, and so is our system. Assuming that the blockchain is secure, the money locked in the blockchain is protected, and hence the payment involved in the system is also safe. We claim that our system takes care of essential security aspects and provides fairness to the parties involved in the system. We state the security properties achieved by our protocol.
	
	\subsection{Fairness}
   We discuss the fairness of each party i.e., Patient, Hospital Authority, Insurance Company, and Research Community. Even if one of the parties acts malicious or more than one party colludes and tries to cheat, the honest party will not lose money or become a victim. In any situation, the malicious parties get penalized or reprimanded. 
   
	\begin{proposition}
   (Patient's Fairness) The honest patient must not lose money or gets treated unfairly, no matter if the other parties (Hospital/Insurance Company) are behaving maliciously (say, they try to cheat the patient in various possible ways), under the assumption that the owner of Medical Data Repository is semi-trusted and the underlying blockchain is secure.
   \end{proposition}
	\textbf{Proof: } We will prove the proposition in the cases where an honest patient's interest might get compromised. Specifically, we will be analyzing the security on the grounds of price, data, and responsiveness.
	\begin{itemize}[leftmargin=*]
	    \item \textbf{Hospital Authority (HA) is malicious}
	            \paragraph*{{\textit{\underline{HA tries to overcharge after treatment}}}} Initially, at the time of admission, the Hospital Authority examines the medical history of the patient and generates a bill with an estimated cost, and it gets recorded in the $BC$. If the patient agrees with it, then only he or she would proceed further to take the services from the hospital. The final cost of check-up must not be higher than this estimated cost. Therefore, in our system, a patient knows the upper limit of the total check-up cost beforehand. It might be possible that the hospital has overcharged, or the patient is not fully satisfied with the treatment, then both the parties can negotiate on this bill and come to a joint agreement (however, it can never be higher than the estimated cost, that was given initially).
	            
	            \paragraph*{{\textit{\underline{HA sends wrong medical files to patient}}}} Medical Data or EHR (Electronic Health Record) is a crucial factor in determining the fairness of the protocol. The patients have to be ensured of receiving the correct file and if not, then there is a provision of raising a complaint and aborting from the protocol. Before actually sending the medical reports to the patient, the hospital first splits the file into multiple chunks of equal size. The Merkle root of the file chunks is obtained; the hospital keeps this Merkle root and the signature on it in the blockchain. Let this root be $M_1$. The hospital encrypts the entire Merkle tree's input, starting from  the leaf level, which is the chunks of the file, to the intermediate level up to the root, using a key and then computes the Merkle root of the encrypted inputs. Let this root be $M_2$. This Merkle root is concatenated with the patient identifier and the date before being signed by the hospital. The encrypted file Merkle root, along with the digital signature, is put in the immutable blockchain (refer to $<M_2, S_2>$ in Figure~\ref{fig:Image3}). Next, the hospital sends the encrypted inputs to the patient using some fair exchange protocol (in our case, we are using fairswap protocol \cite{dziembowski2018fairswap}). Upon receiving the encrypted inputs, the patient re-computes the Merkle root, verifies the result with the stored root in the blockchain, and verifies the signatures to ensure that the file indeed came from the same hospital (proof of authenticity).\newline
                Date and patient identifier has been used along with $M_2$ for the signature to ensure that the hospital is held accountable if in case it sends encrypted inputs of a file belonging to a different person, or it sends a patient’s previous medical file. If in case the hospital keeps the date and patient identifier correct with a mismatched file, the patient on deriving the file after the revelation of the key might come to that file is corrupted. The Merkle root of the file and the signature on it that was kept in the blockchain serves as a sufficient means to hold the hospital for its maliciousness.\newline
                The key closely coupled with the file process also needs to be rightly served to the patient to ensure a fair protocol. There are two ways in which the key can become an issue for the patient. In the first case, the hospital keeps the commitment of one key in the blockchain and reveals a different key to be used for decryption. In the second case, right after encrypting with a key, the hospital commits to a different key and reveals the same. In this case, the patient will decrypt the file with the second key and on finding a mismatch in the gate output computation, it will lodge a complaint which will be resolved on-chain by penalizing the malicious party.
                
                \paragraph*{{\textit{\underline{HA repudiates}}}} In case the hospital authority provides a wrong file and computes the digital fingerprint as per the wrongly generated file, then the contract cannot detect whether things are out of place. When the patient later detects it has received the wrong medical report, in such a case the transcript of the interaction between hospital and patient present in blockchain can be used for accusing the authorities and taking legal action if needed. 
	            
	            \paragraph*{{\textit{\underline{HA becomes unresponsive in the midst of the protocol}}}}
	            A timely response is necessary for a fair protocol. A party can stop responding, which can lead to an indefinite delay in the termination of the protocol. So, when the patient locks the estimated bill amount in the smart contract; the hospital needs to start the treatment within a fixed time window (a tunable parameter that can be set as per the system requirement). Hospital immediately invokes a function $startTreatment$ (defined in SC\_P\_HA) and updates the state. That essentially makes a transaction entry in the blockchain. Failure to start treatment within a limited period causes the patient to unlock his/her money. A patient can then seek treatment from some other hospitals. We can also penalize the hospital for this negligence (by deducting some amount of locked estimated cost of the hospital). That is the reason behind locking the estimated cost amount by the hospital. If the hospital becomes unresponsive at any point in the protocol, the patient can call the respective exit function to abort from the protocol, and we can also penalize a hospital for any such malpractice by deducting its locked money.
	            \item \textbf{Insurance Company (IC) is malicious}
	        
                \paragraph*{{\textit{\underline{IC tries to cheat user/policy buyer by not approving claim}}}}
                The hash of the terms \& conditions file of the policy is stored in the blockchain, and the patient also verifies the same at the time of buying the policy. Therefore, later on, the insurer can not deny it. Another thing is that when a policy buyer claims his or her medical bill amount, a transaction is made to reflect the same, and hence, the details go to the blockchain, and it must be settled within some fixed time window after the claim. Otherwise, the insurer will be penalized, and in a distributed setup like blockchain, the insurer loses its reputation among the community, which they obviously try to avoid.
                If IC approves a claim partially and the user/claimer is not satisfied with the approval, he/she can take help of the law complaining against the IC. The judge can easily sort out the dispute with the aid of blockchain (as the hash of terms and condition file is kept in BC) 
                In case the IC does not respond to the user's claim at all, then the system will be able to return the price paid upon purchase of the policy to the user. If it is not possible to return the money (due to insufficient security money of IC), the IC would be blacklisted from the system.
                
                \paragraph*{{\textit{\underline{IC steals patient's medical data}}}}
                If we observe the claim process in Figure~\ref{fig:Image4}, the policy buyer needs to grant permission to IC to access his or her data stored in the medical database, so that insurer can verify the authenticity of the claim raised by the policy buyer in the future. However, if DBO releases the medical report, then the Insurance Company can take the data and stop any sort of interaction with the buyer. To prevent this, the Database owner sends the encrypted data (enc\_data in Figure~\ref{fig:Image4}) to the Insurance Company. Encryption is done using a temporary key $K$,  generated by the user itself. This key is used for one time only. Upon receiving the encrypted data, the Insurance Company locks the claimed amount in the smart contract and asks for the key ($K$) from the user. The user would not reveal the key unless the insurer locks the money in the contract. Another thing is that we have discussed about logical data partitioning - a patient/user data can be divided into three different categories - Personal Information, Medical Reports, and Medical Expenses. If privacy is needed, users can only grant access to the Medical Expenses section of the data and hide the other parts. Hence, it takes care of the data privacy aspects as well.
                
                \paragraph*{{\textit{\underline{IC becomes unresponsive in the midst of the protocol}}}}
                For every subsequent function call in the smart contract, there is a fixed time window. If the IC fails to respond on time, then the user can take suitable actions accordingly, e.g., a user can unlock the money for buying a policy if the IC does not reply in Phase two of policy buying. Similarly, at the time of claiming as well, the user can take appropriate measures so that he/she would not lose money, in case the IC becomes unresponsive.
            	\item \textbf{Both Hospital Authority and Insurance Company are malicious} Irrelevant since both the entities (HA and IC) act independently of the patient or policy buyer. The activity of HA does not influence the IC in any way and vice-versa.
	\end{itemize}

	\begin{proposition}
	   (Hospital's Fairness) The Hospital Authority would be able to get its money for all the valid services provided to the patient, in spite of the patient's misbehavior (say, the patient tries to take services from the hospital without paying the bill amount and then leave), under the assumption that the underlying BC is secure.
	\end{proposition}
	\textbf{Proof:} If a patient is malicious, he or she may try to cheat in the following ways:
	
	\paragraph*{{\textit{\underline{Patient does not pay for the medical service}}}}
	    A patient can not deny paying the hospital after taking medical services. Because at the time of taking admission to the hospital, the hospital authority would guess the expected cost of the treatment. The patient is supposed to lock this amount in the smart contract, and only then the treatment begins. Again, at the time of discharge, the hospital authority calculates the exact/final treatment cost. In our case, we have considered that the final cost should not be greater than the estimated cost. Once the patient gives consent over the final bill by putting a signature, then the patient receives all the medical files (i.e., Prescriptions, reports, etc.). Otherwise, the patient does not obtain these medical files (Patient can have the encrypted files, but the decryption key is still with the hospital. It is only revealed after the final bill has been locked in the contract by the patient).
	    
	    \paragraph*{{\textit{\underline{Patient produces a wrong complaint}}}}
	    Receiving the correct file, the patient might try to raise a false complaint to exonerate money from the hospital without paying any money for the obtained services. The encrypted Merkle root ensures that the complaint should be part of the Merkle tree. This is used for checking the validity of the complaint and the fact that the patient is lying gets detected.
	    
    	\paragraph*{{\textit{\underline{Patient becomes unresponsive in the middle of protocol}}}}
	    Different exit points prevent an honest party from waiting indefinitely and keep its money locked in the smart contract. If this happens at any time in the middle of the protocol, the hospital can exit and unlock the money locked in the contract. 

	\begin{proposition}
	   (Insurance Company's Fairness) When a user buys a policy from an Insurance Company, the Insurance Company would be able to get the money stated in the terms while purchasing the policy. There should not be any such instances where a malicious user/buyer enjoys the benefits of a policy without paying the insurance company's policy price. Another aspect is that a user or policy buyer can not make a false claim and cheat the Insurer and force them to pay money.
	\end{proposition}
	\textbf{Proof:} If User (Policy buyer) is malicious, he or she may try to cheat in the following ways:
    	\paragraph*{{\textit{\underline{User gets benefited from a policy without paying for it.}}}}
	    In the first phase of policy buying (Algorithm~\ref{algo:Algo5}), the policy buyer expresses his or her interest in keeping the money (i.e., Policy Price) locked in the contract. The user also puts the hash of the terms and conditions files into the BC. The IC verifies the same in the second phase of buying. If everything is found right, then the smart contract transfers the policy price to the IC's account, and a unique policy ID is communicated to the buyer. So, it is not possible to have a policy without paying money for the same.
	    
	    \paragraph*{{\textit{\underline{User tries to generate a false claim}}}}
	    When the user claims reimbursement for his/her medical expenses (Algorithm~\ref{algo:Algo6}), the system checks whether the claim has already been sanctioned. Then the system checks whether all files related to medical expenses are correct (by matching their hash and by verifying the signature of Hospital Authority and patient, stored in the BC). Finally, the system checks whether the claim follows the terms and conditions of the policy. Once it is verified, the claim is sanctioned (partially or full). So, the Insurance Company's fairness is preserved. Users cannot cheat an Insurance Company in our system.
	    
	    \paragraph*{{\textit{\underline{User becomes unresponsive in the middle of the protocol}}}}
	    There is no motivation as such for the user to become unresponsive (as it is the user and not the IC, whose money has been locked in the system). However, a user may be unresponsive in the claiming process. After making an insurance claim, the user might delay revealing the key when the IC needs the key for decryption of the encrypted files. If it happens, then IC can exit from the protocol without approving any money claimed by the user and unlocks the IC's money locked in the contract. Therefore, IC does not get cheated in case the user becomes unresponsive.

	\begin{proposition}
	   (Research Community's Fairness) In our system, the Research Community(RC) should have sufficient confidence regarding the correctness of the medical data, which they received from a medical repository.
	\end{proposition}
	\textbf{Proof: }
	It is essential to provide correct medical data to the RC, as analysis of wrong data may lead to an incorrect outcome in research, which is not desirable for advanced medical science. At the same time, it is not good practice to reveal the patient's identity while sending the data to RC (i.e., preserving patient privacy). Keeping these facts in mind, when RC requests for medical data, the DBO (which is a semi-trusted party appointed by the Government) sends the aggregated medical data of the specific disease (without disclosing the patients' identity) to the RC. It is right that we are dependent on the trust assumption of the DBO here. But DBO is not trusted. Whatever DBO sends to the RC, the hash of it as well as the signature over the computed hash, is kept in the BC. Hence, the DBO is also accountable for the activities, which he or she does. Once the data is received, the RC computes the hash of it and verifies the signature. If everything is found right, collecting the medical data, the RC can carry on its research work with sufficient confidence. 
	
	\subsection{Privacy}
	A patient's medical data is sensitive information. If a person's health record is available publicly, he or she may face embarrassment and might be subjected to discrimination in daily life. Hence, it must be ensured that access to patient data is provided only with the patient's consent. 
	
	\begin{proposition}
	   (Patient's Privacy) In our proposed system, none of the entities can access a patient's data unless it has been granted permission. At the same time, personally identifiable attributes o the patient remains hidden from public view.
	\end{proposition}
	\textbf{Proof: }
	 A patient does not reveal the attributes like name, age, address, contact number, etc while registering on the blockchain. Instead, the hash of these attributes is kept in the BC at the time of registration. Since our Healthcare model is patient-centric, no one can see the patient's medical data without his/her consent.  The patient defines his/her access control matrix, mentioning the identities who have access to the patient's data. The patient has the provision to update the matrix from time to time - he or she can grant access permission to another party or revoke access permission from a party. The access control matrix is stored in the BC.
	 
	\subsection{Data Security}
	Tampering a patient's medical data might have catastrophic consequences. If medicine doses are changed or if the ailment diagnosed is modified, then this could lead to life risk. Also, a patient may try to tamper with his or her medical history to claim reimbursement from the insurance company illegitimately. We discuss how our protocol mitigates such problems.
	
	\begin{proposition}
	   Proposition 6 (Data Security) In our proposed system, no one, including the patient, can tamper with the medical data and misuse it for malicious purposes.
	\end{proposition}
	\textbf{Proof: } Since we have built our protocol using blockchain as the underlying framework, the security of medical data follows from the inherent property of blockchain’s immutability.
	
	\section{Performance Evaluation and Discussion}
	\subsection{Implementation Setup}
	
	We have implemented the Health Care Management System on Ethereum test networks in a system having Intel(R) Core(TM) i7-6700HQ running Linux Mint 18.04 19.1 (Tessa), 64-bit operating system using 16.00GiB of RAM. We have used the Ropsten test network and an infura endpoint.

	
	
	\subsection{Result}
	The main objective was to get the various results from the implementation. The two main obvious factors that determine the feasibility of any blockchain model are cost and time.
    Since the implementation is in the Ethereum blockchain, we have taken the transaction cost (Gas Cost converted to equivalent Dollar) of the protocols and the time taken to execute the same. Particularly, the entire implementation has been deployed and run in the Ropsten test network. The time as a metric when taken from the public network like Ropsten will give a much more pragmatic insight compared to the private network setup.
    
    \begin{table}[h!]
		\centering
		\caption{Deployment Addresses of Smart Contracts}
		\label{tab:TableDepAdd}
		\scalebox{0.8}{
			\begin{tabular}{c|c}
				\hline \hline
				\textbf{Smart Contract} & \textbf{Address} \\ \hline \hline
				SC\_Registration & 0x5a818296705cC24Feec4CfEAF1DfdaE056fEf037 \\ \hline
				{SC\_P\_HA\_1} & 0x9528dA5753ae928Eb1e0284C7b1771e2FC17a766\\ \hline
				{SC\_P\_HA\_2} & 0x7b88e153aC1b2BCA865CD58E1082f50Ed69f4c3c\\ \hline
				{SC\_P\_DBO} & 0xC062E1eF5EdB815bcF5B93C6BaD497ABCA407f31\\ \hline
				{SC\_P\_IC\_DBO} & 0xD519535972d006DD72AbBd60453Ae78747065B5e \\ \hline
				{SC\_RC\_DBO} & 0xafb22029C41cA8a4c04D0a07284d2b2c257aE723\\ \hline
			\end{tabular}
		}
	\end{table}
	
	\begin{table}[h!]
		\centering
		\caption{Deployment Cost of Smart Contracts}
		\label{tab:TableDepCost}
		\scalebox{0.8}{
			\begin{tabular}{c|c}
				\hline \hline
				\textbf{Smart Contract} & \textbf{Deployment Cost(Ether)} \\ \hline \hline
				SC\_Registration & 
				0.0353157\\ \hline
				{SC\_P\_HA\_1} & 
				0.0994202\\ \hline
				{SC\_P\_HA\_2} & 
				0.0583187\\ \hline
				{SC\_P\_DBO} & 
				0.04783231\\ \hline
				{SC\_P\_IC\_DBO} & 
				0.0516384 \\ \hline
				{SC\_RC\_DBO} & 
				0.0113837\\ \hline
			\end{tabular}%
		}
	\end{table}

\begin{figure} [!ht]
		\begin{center}  
			\includegraphics[width=3.5in, height=1.9in]{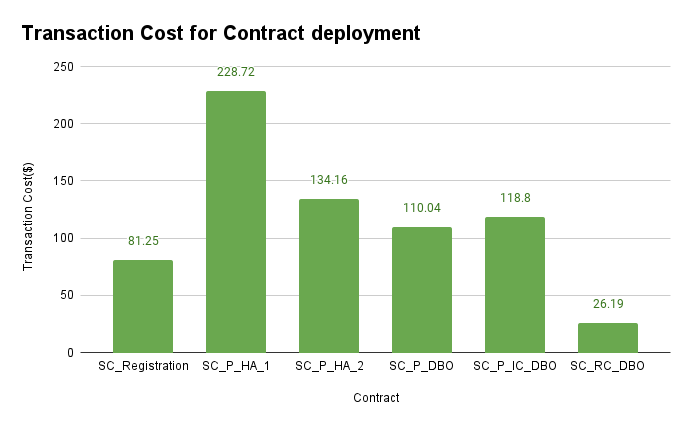}
			\setlength{\belowcaptionskip}{-15 pt}
			\caption{\small \sl Transaction Cost for Contract Deployment}
			\label{fig:Image7}
		\end{center}  
	\end{figure}

\begin{figure} [!ht]
		\begin{center}  
			\includegraphics[width=3.5in, height=1.9in]{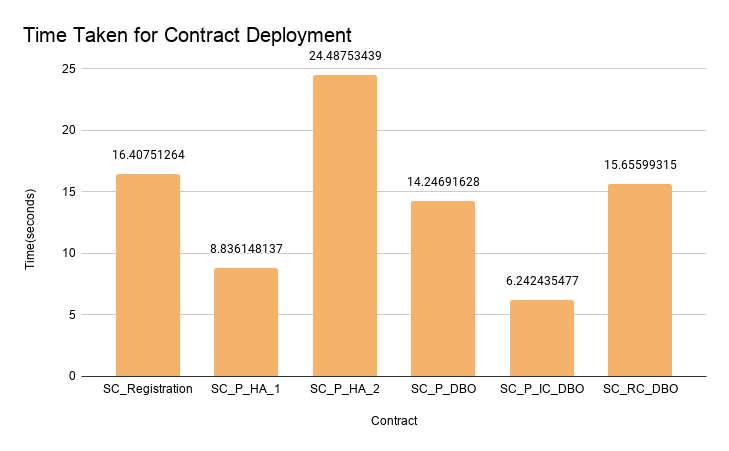}
			\setlength{\belowcaptionskip}{-15 pt}
			\caption{\small \sl Time for Contract Deployment}
			\label{fig:Image8}
		\end{center}  
	\end{figure}

    Table~\ref{tab:TableDepAdd} specifies the addresses of the deployed contracts. The contract deployment is a one-time occurrence. The transaction cost and the time taken for each contract deployment have been depicted in Figure~\ref{fig:Image7} and Figure~\ref{fig:Image8} respectively. The gas price was 18.9 Gwei and the ether cost was 2300.54 dollars at the time of deployment. Depending upon the size of the contracts, the deployment cost varies (Table~\ref{tab:TableDepCost}). These are one-time costs. So, once deployed, we can get the benefits throughout the usage of this protocol.
    
    The smart contracts for patients and hospitals (SC\_P\_HA\_1  and SC\_P\_HA\_2) have been split into two parts, citing the limited gas limit for blocks in Ethereum. The high gas for the collective patient and hospital contracts is a reflection of the slightly higher steps involved in the protocol. 
    
	\begin{figure} [!ht]
		\begin{center}  
			\includegraphics[width=3.5in, height=1.9in]{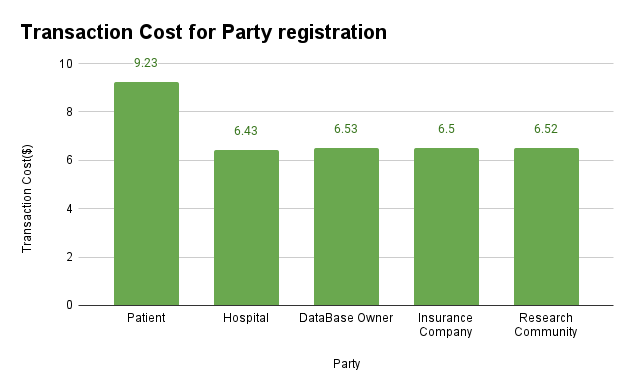}
			\setlength{\belowcaptionskip}{-15 pt}
			\caption{\small \sl Transaction Cost for Party Registration}
			\label{fig:Image9}
		\end{center}  
	\end{figure}
	
    \begin{figure} [!ht]
		\begin{center}  
			\includegraphics[width=3.5in, height=1.9in]{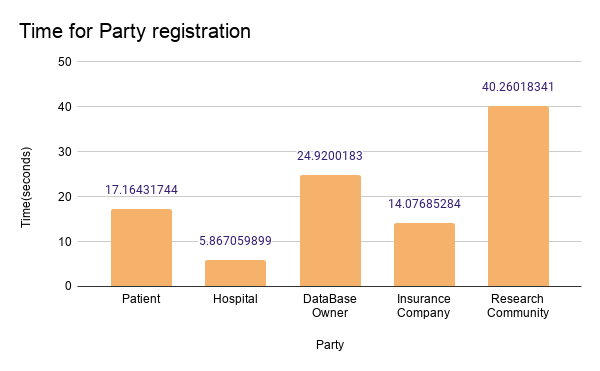}
			\setlength{\belowcaptionskip}{-15 pt}
			\caption{\small \sl Time for Party Registration}
			\label{fig:Image10}
		\end{center}  
	\end{figure}

    \begin{figure} [!ht]
		\begin{center}  
			\includegraphics[width=3.5in, height=1.9in]{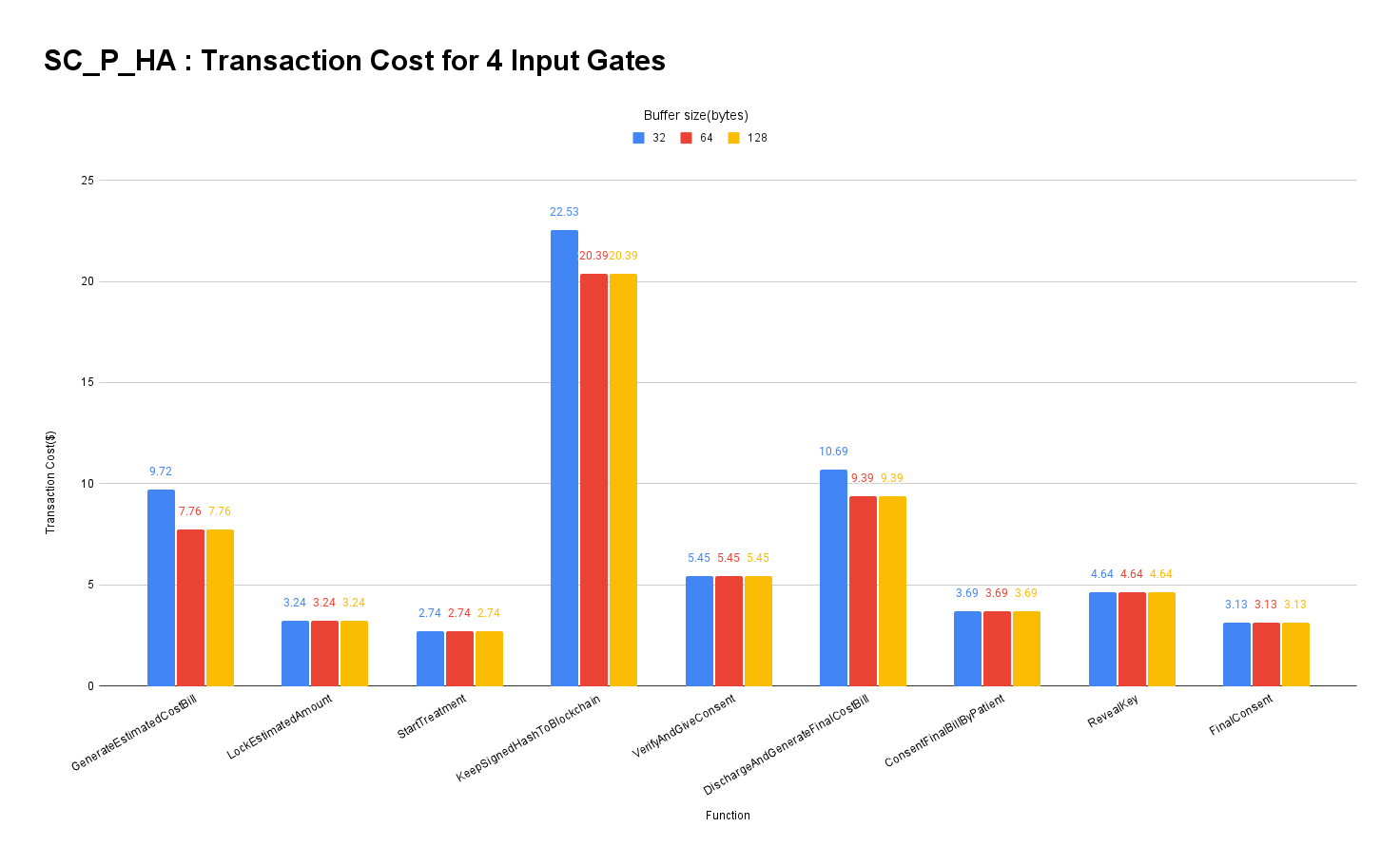}
			\setlength{\belowcaptionskip}{-15 pt}
			\caption{\small \sl {SC\_P\_HA} Transaction Cost for 4 Input Gates
				\label{fig:Image11}}
		\end{center}  
	\end{figure}
	
    \begin{figure} [!ht]
		\begin{center}  
			\includegraphics[width=3.5in, height=1.9in]{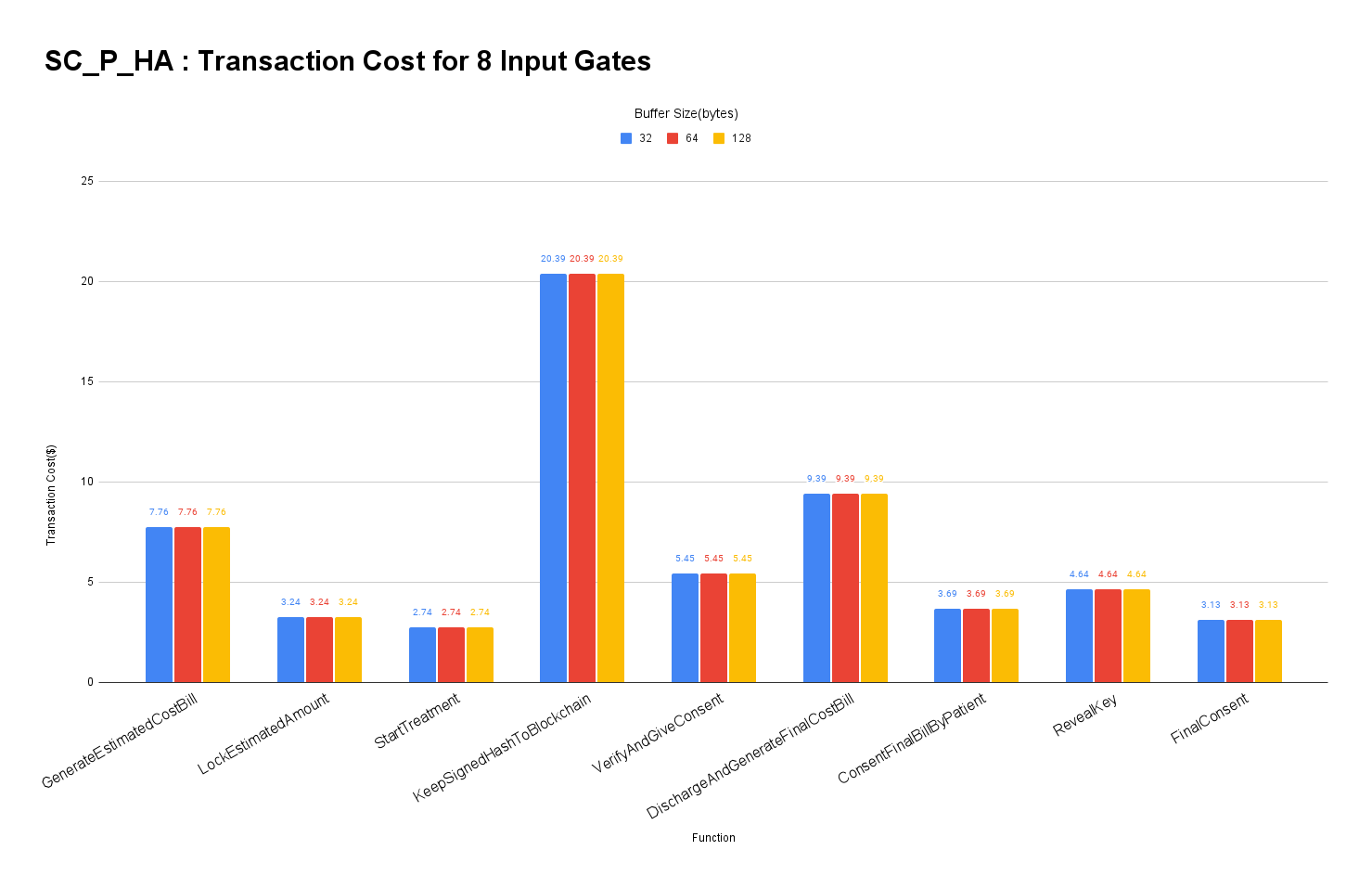}
			\setlength{\belowcaptionskip}{-15 pt}
			\caption{\small \sl {SC\_P\_HA} Transaction Cost for 8 Input Gates
				\label{fig:Image12}}
		\end{center}  
	\end{figure}

	

	

    \begin{figure} [!ht]
		\begin{center}  
			\includegraphics[width=3.5in, height=1.9in]{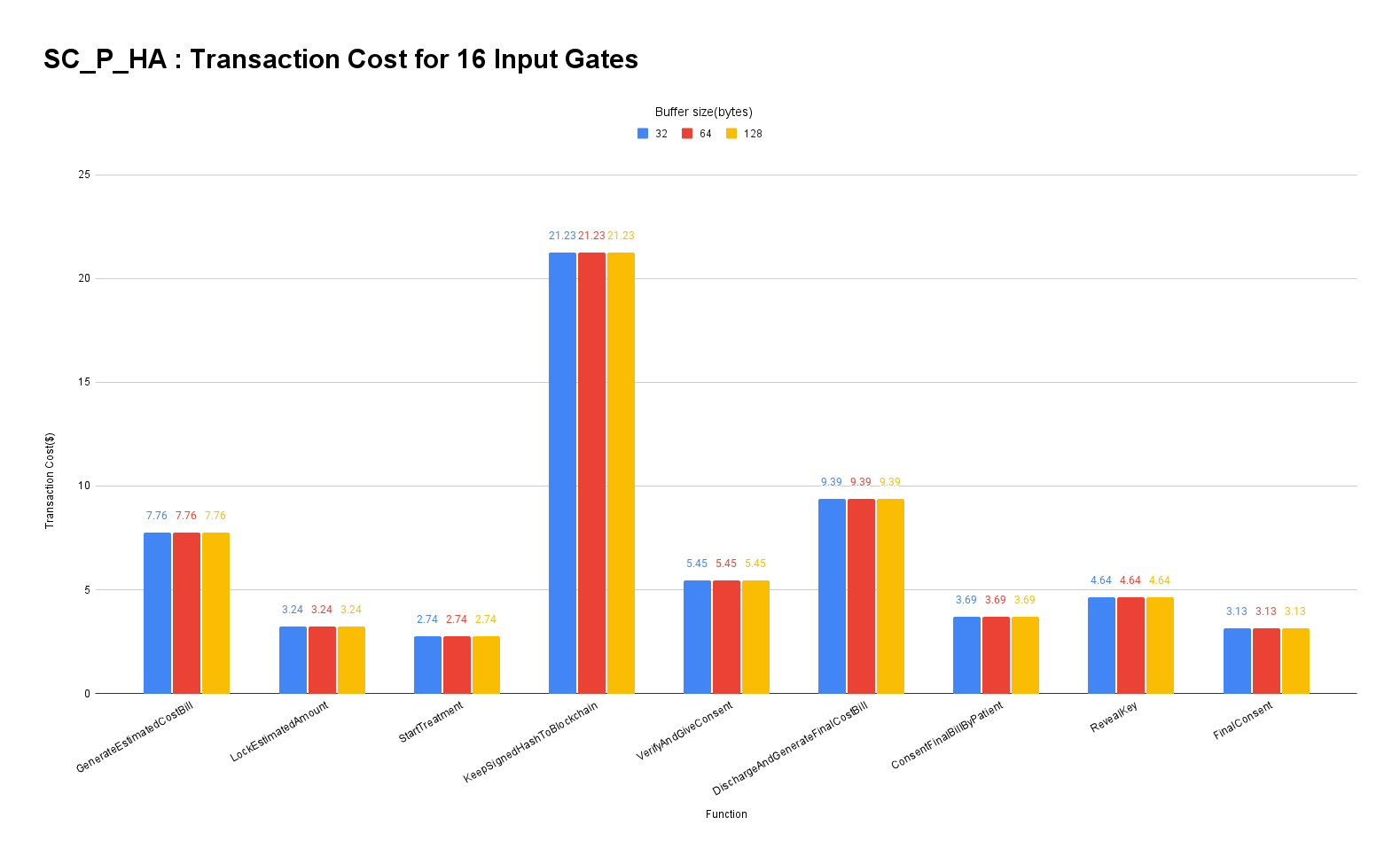}
			\setlength{\belowcaptionskip}{-15 pt}
			\caption{\small \sl {SC\_P\_HA} Transaction Cost for 16 Input Gates
				\label{fig:Image13}}
		\end{center}  
	\end{figure}
	

    \begin{figure} [!ht]
		\begin{center}  
			\includegraphics[width=3.5in, height=1.9in]{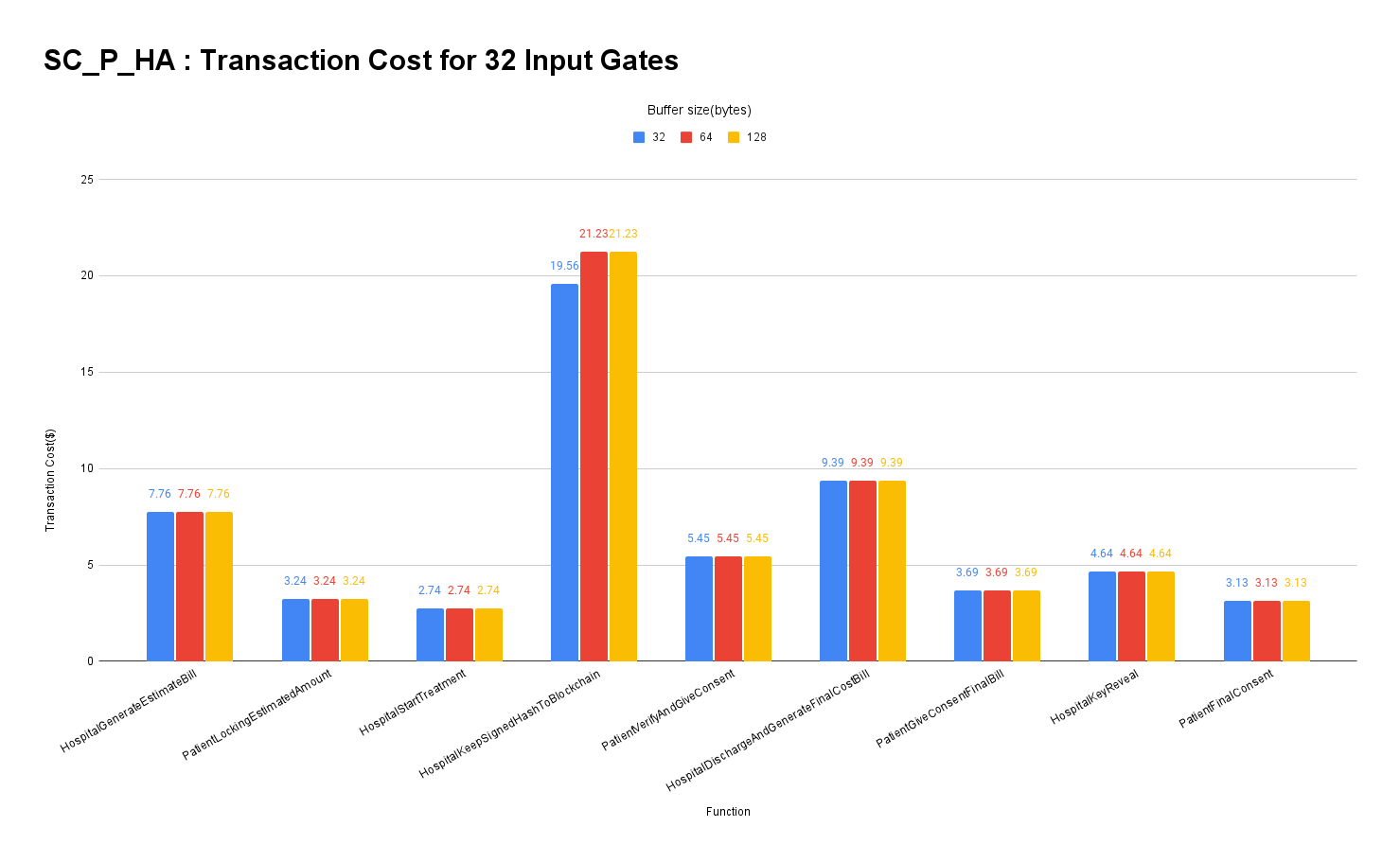}
			\setlength{\belowcaptionskip}{-15 pt}
			\caption{\small \sl {SC\_P\_HA} Transaction Cost for 32 Input Gates
				\label{fig:Image14}}
		\end{center}  
	\end{figure}
	
	
	\begin{figure} [!ht]
		\begin{center}  
			\includegraphics[width=3.5in, height=1.9in]{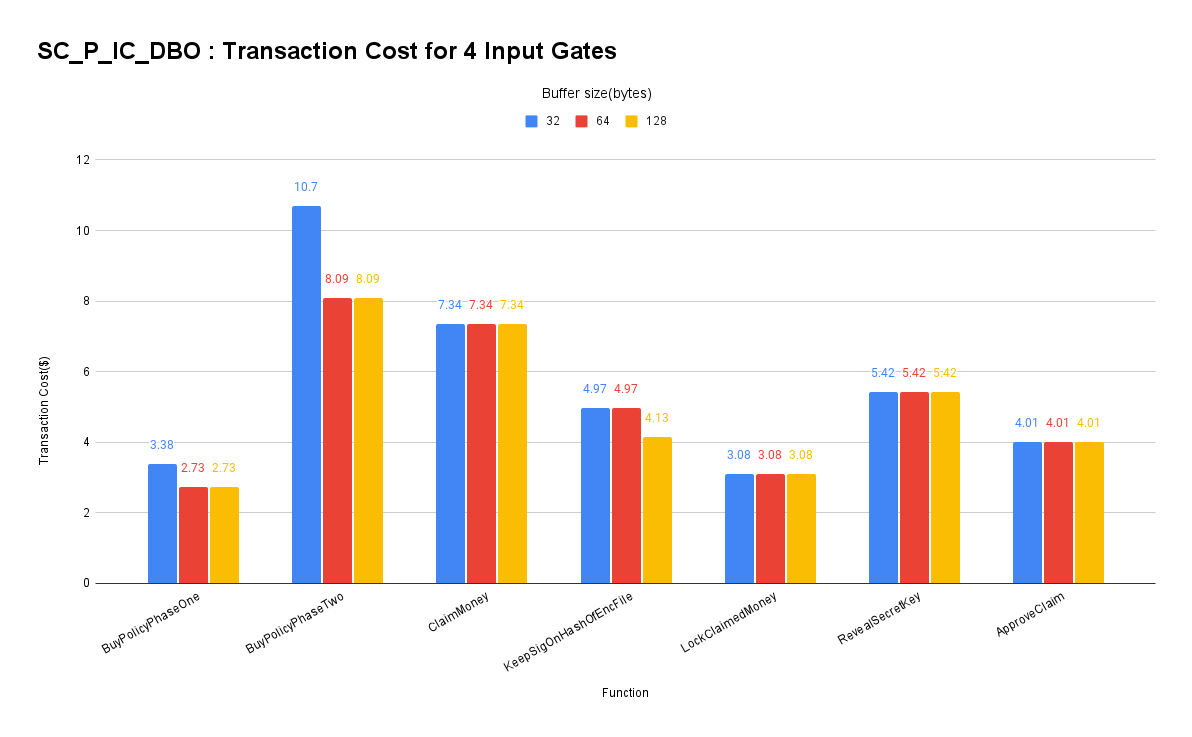}
			\setlength{\belowcaptionskip}{-15 pt}
			\caption{\small \sl {SC\_P\_IC\_DBO} Transaction Cost for 4 Input Gates
				\label{fig:Image15}}
		\end{center}  
	\end{figure}
	
	
	\begin{figure} [!ht]
		\begin{center}  
			\includegraphics[width=3.5in, height=1.9in]{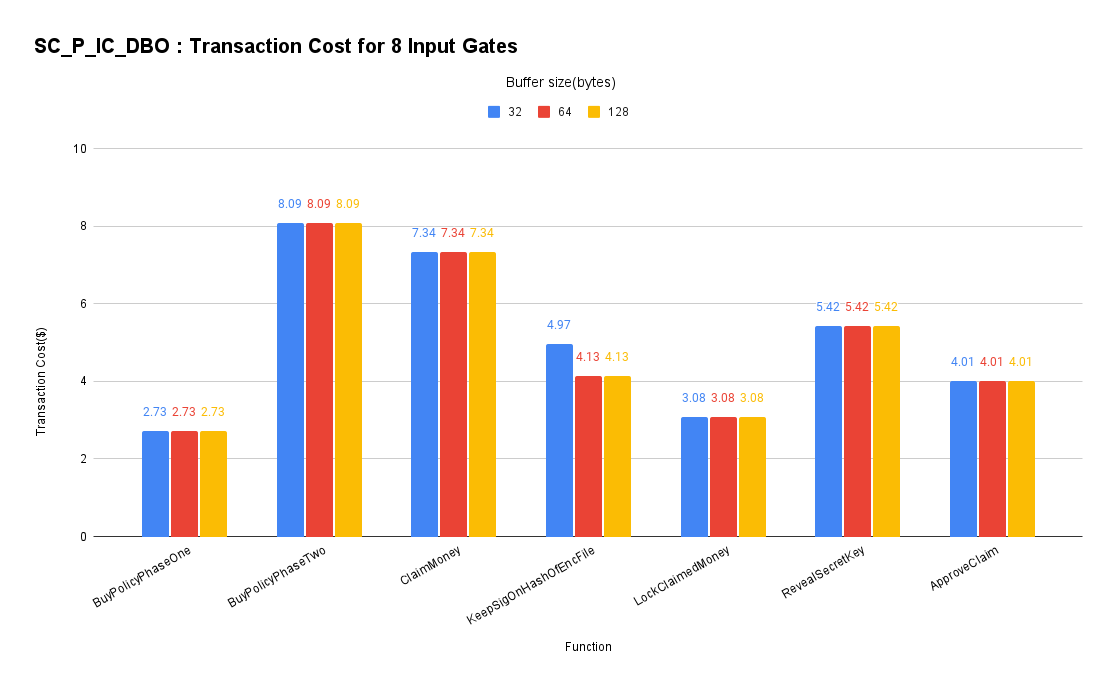}
			\setlength{\belowcaptionskip}{-15 pt}
			\caption{\small \sl {SC\_P\_IC\_DBO} Transaction Cost for 8 Input Gates
				\label{fig:Image16}}
		\end{center}  
	\end{figure}
	

    \begin{figure} [!ht]
		\begin{center}  
			\includegraphics[width=3.5in, height=1.9in]{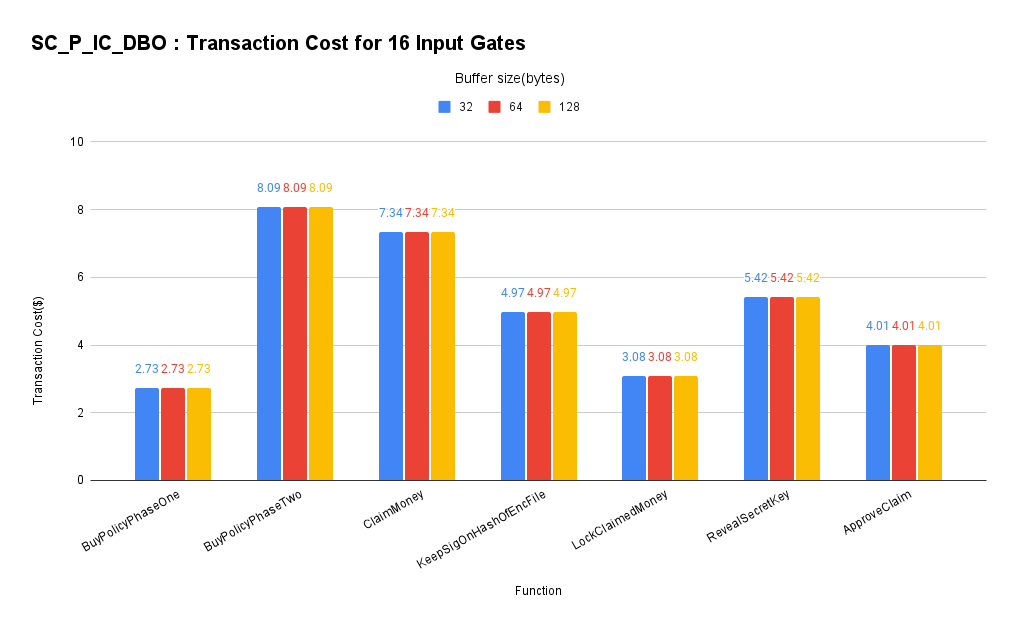}
			\setlength{\belowcaptionskip}{-15 pt}
			\caption{\small \sl {SC\_P\_IC\_DBO} Transaction Cost for 16 Input Gates
				\label{fig:Image17}}
		\end{center}  
	\end{figure}
	
	
	\begin{figure} [!ht]
		\begin{center}  
			\includegraphics[width=3.5in, height=1.9in]{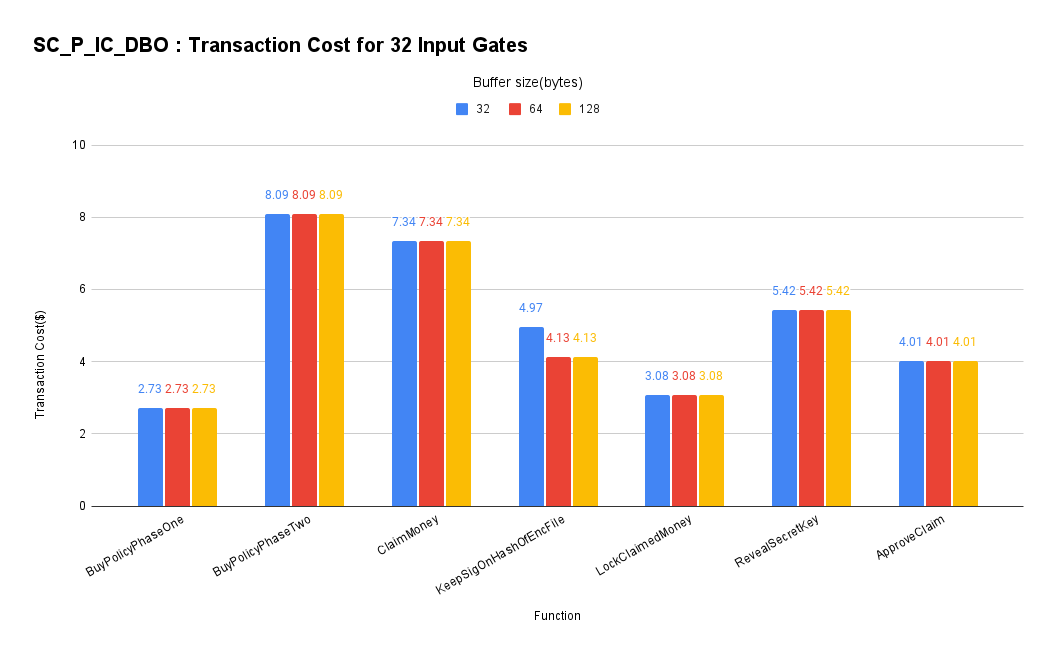}
			\setlength{\belowcaptionskip}{-15 pt}
			\caption{\small \sl {SC\_P\_IC\_DBO} Transaction Cost for 32 Input Gates
				\label{fig:Image18}}
		\end{center}  
	\end{figure}
	

	
	\begin{figure} [!ht]
		\begin{center}  
			\includegraphics[width=3.5in, height=1.9in]{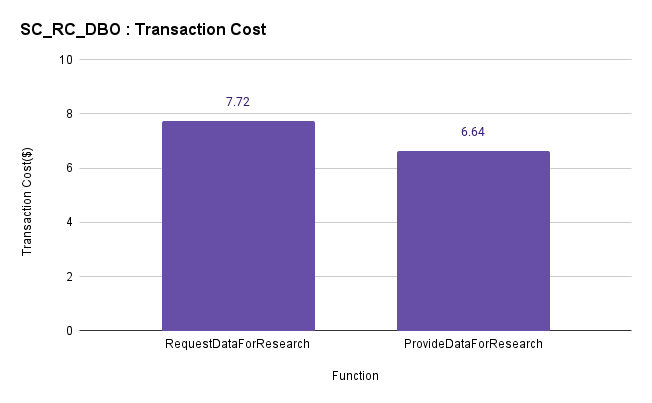}
			\setlength{\belowcaptionskip}{-15 pt}
			\caption{\small \sl {SC\_RC\_DBO} Transaction Cost
				\label{fig:Image19}}
		\end{center}  
	\end{figure}

	

    The transaction costs and time taken associated with the entities’ registration process, depicted in Figure~\ref{fig:Image9} and Figure~\ref{fig:Image10} respectively, are similar in scale for different parties except for the patients which are slightly higher due to the few more variables involved in the registration for patients.

    The transaction cost for many of the functions involved in the protocol depends on the size of the files. The file size may vary depending upon the treatment and the corresponding result produced. The base file is constructed indifferently for subsequent usage in the protocol. The file is divided into numbers of chunks. The number and size of the chunks are the varying parameters, which are referred to as the number of input gates and the buffer size of the gate respectively in the paper. 

    We have shown the cost associated with various functions call for the contracts SC\_P\_HA (Figure~\ref{fig:Image11}, \ref{fig:Image12}, \ref{fig:Image13}, \ref{fig:Image14}) and SC\_P\_IC\_DBO (Figure~\ref{fig:Image15}, \ref{fig:Image16}, \ref{fig:Image17}, \ref{fig:Image18}) with varying number of input gates and the buffer size. The number of input gates is varied in the range of 4, 8, 16, 32 and the buffer sizes used are 32, 64, and 128. The file size can be derived by multiplying the number of input gates and the buffer size. So, a file having 4 input gates and a buffer size of 32 would have a file size of 128 bytes.
    
     We find from the graphs of SC\_P\_HA  (Figure~\ref{fig:Image11}, \ref{fig:Image12}, \ref{fig:Image13}, \ref{fig:Image14}) that they follow a similar kind of trend with varying numbers of input gates. While keeping the number of input gates constant and varying the buffer size, the transaction cost for different functions hardly varies. 
     Also, this holds true for the graphs SC\_P\_IC\_DBO (Figure~\ref{fig:Image15}, \ref{fig:Image16}, \ref{fig:Image17}, \ref{fig:Image18}) as well. The cost associated with the functions of SC\_RC\_DBO is shown in Figure~\ref{fig:Image19}.
     As shown in the graphs, some functions require a higher cost. This is due to their heavy functionalities. It is viable as the utilities of the functions outweigh the cost.
     
	\section{Conclusion}
	The use of blockchain in healthcare systems plays an important role in the healthcare domain. It can result in automated data collection and verification processes,  enables the collection of correct and aggregated data from various sources that are immutable, tamper-resistant, and provide secured data, with a reduced probability of cybercrime \cite{tanwar2020blockchain}. It also supports distributed data, with redundancy and fault tolerance of the system. Being a distributed system, it also eliminates the role of a trusted third party. In this paper, we have depicted a novel \emph{Secure and Smart Healthcare System} where every involved party's fairness is preserved without trusting each other. Electronic health records (EHR) are tamper-proof and free from any unauthorized access in our healthcare system enabled by Blockchain Technology. This shows the blockchain capability and importance in various areas and proves that it could be the next revolutionary technology for replacing current healthcare systems. 
	
	Medical data is of utmost importance for research purposes. Nowadays, it is challenging to get fresh unaltered medical data from the source, which necessarily leads to poor research results while applying various data mining techniques. Cleaning raw data is one of the hardest jobs. We have also tried to focus on this area so that Research Community can get the medical data with a certain level of confidence about data integrity.  We have proposed, prototyped, and deployed our Healthcare system, which works fine in the private network as well as in the Ropsten test network. We have built the system in the permissionless setting of the blockchain environment. We have ensured the patient's privacy does not get compromised. Experimental result shows the satisfactory outcome of various performance metrics. Further, we can improve the performance of this model by making further optimizations that are left as a part of our future work. 
	
	\bibliographystyle{ieeetr}
	\bibliography{mybibliography}

\begin{thebibliography}{10}

\bibitem{reyna2018blockchain}
A.~Reyna, C.~Mart{\'\i}n, J.~Chen, E.~Soler, and M.~D{\'\i}az, ``On blockchain
  and its integration with iot. challenges and opportunities,'' {\em Future
  generation computer systems}, vol.~88, pp.~173--190, 2018.

\bibitem{dorri2017blockchain}
A.~Dorri, S.~S. Kanhere, R.~Jurdak, and P.~Gauravaram, ``Blockchain for iot
  security and privacy: The case study of a smart home,'' in {\em 2017 IEEE
  international conference on pervasive computing and communications workshops
  (PerCom workshops)}, pp.~618--623, IEEE, 2017.

\bibitem{nakamoto2008re}
S.~Nakamoto, ``Re: Bitcoin p2p e-cash paper,'' {\em The Cryptography Mailing
  List}, 2008.

\bibitem{zheng2017overview}
Z.~Zheng, S.~Xie, H.~Dai, X.~Chen, and H.~Wang, ``An overview of blockchain
  technology: Architecture, consensus, and future trends,'' in {\em 2017 IEEE
  international congress on big data (BigData congress)}, pp.~557--564, IEEE,
  2017.

\bibitem{zheng2018blockchain}
Z.~Zheng, S.~Xie, H.-N. Dai, X.~Chen, and H.~Wang, ``Blockchain challenges and
  opportunities: A survey,'' {\em International Journal of Web and Grid
  Services}, vol.~14, no.~4, pp.~352--375, 2018.

\bibitem{buterin2014next}
V.~Buterin {\em et~al.}, ``A next-generation smart contract and decentralized
  application platform,'' {\em white paper}, vol.~3, no.~37, 2014.

\bibitem{MedShare}
Q.~Xia, E.~Sifah, K.~Asamoah, J.~Gao, X.~Du, and M.~Guizani, ``Medshare:
  Trust-less medical data sharing among cloud service providers via
  blockchain,'' {\em IEEE Access}, vol.~PP, pp.~1--1, 07 2017.

\bibitem{ContinuousPatientMonitoring}
M.~A. {Uddin}, A.~{Stranieri}, I.~{Gondal}, and V.~{Balasubramanian},
  ``Continuous patient monitoring with a patient centric agent: A block
  architecture,'' {\em IEEE Access}, vol.~6, pp.~32700--32726, 2018.

\bibitem{FHIRChain}
P.~Zhang, J.~White, D.~Schmidt, G.~Lenz, and S.~Rosenbloom, ``Fhirchain:
  Applying blockchain to securely and scalably share clinical data,'' {\em
  Computational and Structural Biotechnology Journal}, vol.~16, 07 2018.

\bibitem{RemotePatientMonitoring}
K.~Griggs, O.~Ossipova, C.~Kohlios, A.~Baccarini, E.~Howson, and T.~Hayajneh,
  ``Healthcare blockchain system using smart contracts for secure automated
  remote patient monitoring,'' {\em Journal of Medical Systems}, vol.~42, 07
  2018.

\bibitem{DataPreservationSystem}
H.~Li, L.~Zhu, M.~Shen, F.~Gao, X.~Tao, and S.~Liu, ``Blockchain-based data
  preservation system for medical data,'' {\em Journal of Medical Systems},
  vol.~42, 08 2018.

\bibitem{AccessControl}
J.~Dias, L.~Reis, H.~Ferreira, and A.~Martins, ``Blockchain for access control
  in e-health scenarios,'' {\em arXiv preprint arXiv:1805.12267}, 05 2018.

\bibitem{DecentralizedAccountability}
X.~Liang, S.~Shetty, J.~Zhao, D.~Bowden, D.~Li, and J.~Liu, ``Towards
  decentralized accountability and self-sovereignty in healthcare systems,'' in
  {\em International conference on information and communications security},
  pp.~387--398, Springer, 2017.

\bibitem{yue2016healthcare}
X.~Yue, H.~Wang, D.~Jin, M.~Li, and W.~Jiang, ``Healthcare data gateways: found
  healthcare intelligence on blockchain with novel privacy risk control,'' {\em
  Journal of medical systems}, vol.~40, no.~10, pp.~1--8, 2016.

\bibitem{zhang2016secure}
J.~Zhang, N.~Xue, and X.~Huang, ``A secure system for pervasive social
  network-based healthcare,'' {\em Ieee Access}, vol.~4, pp.~9239--9250, 2016.

\bibitem{liang2017integrating}
X.~Liang, J.~Zhao, S.~Shetty, J.~Liu, and D.~Li, ``Integrating blockchain for
  data sharing and collaboration in mobile healthcare applications,'' in {\em
  2017 IEEE 28th annual international symposium on personal, indoor, and mobile
  radio communications (PIMRC)}, pp.~1--5, IEEE, 2017.

\bibitem{jiang2018blochie}
S.~Jiang, J.~Cao, H.~Wu, Y.~Yang, M.~Ma, and J.~He, ``Blochie: a
  blockchain-based platform for healthcare information exchange,'' in {\em 2018
  ieee international conference on smart computing (smartcomp)}, pp.~49--56,
  IEEE, 2018.

\bibitem{fan2018medblock}
K.~Fan, S.~Wang, Y.~Ren, H.~Li, and Y.~Yang, ``Medblock: Efficient and secure
  medical data sharing via blockchain,'' {\em Journal of medical systems},
  vol.~42, no.~8, pp.~1--11, 2018.

\bibitem{wang2018secure}
H.~Wang and Y.~Song, ``Secure cloud-based ehr system using attribute-based
  cryptosystem and blockchain,'' {\em Journal of medical systems}, vol.~42,
  no.~8, pp.~1--9, 2018.

\bibitem{guo2018secure}
R.~Guo, H.~Shi, Q.~Zhao, and D.~Zheng, ``Secure attribute-based signature
  scheme with multiple authorities for blockchain in electronic health records
  systems,'' {\em IEEE access}, vol.~6, pp.~11676--11686, 2018.

\bibitem{sun2018decentralizing}
Y.~Sun, R.~Zhang, X.~Wang, K.~Gao, and L.~Liu, ``A decentralizing
  attribute-based signature for healthcare blockchain,'' in {\em 2018 27th
  International conference on computer communication and networks (ICCCN)},
  pp.~1--9, IEEE, 2018.

\bibitem{zhang2018blockchain}
X.~Zhang and S.~Poslad, ``Blockchain support for flexible queries with granular
  access control to electronic medical records (emr),'' in {\em 2018 IEEE
  International conference on communications (ICC)}, pp.~1--6, IEEE, 2018.

\bibitem{yang2018design}
G.~Yang and C.~Li, ``A design of blockchain-based architecture for the security
  of electronic health record (ehr) systems,'' in {\em 2018 IEEE International
  conference on cloud computing technology and science (CloudCom)},
  pp.~261--265, IEEE, 2018.

\bibitem{sukhwani2017performance}
H.~Sukhwani, J.~M. Mart{\'\i}nez, X.~Chang, K.~S. Trivedi, and A.~Rindos,
  ``Performance modeling of pbft consensus process for permissioned blockchain
  network (hyperledger fabric),'' in {\em 2017 IEEE 36th Symposium on Reliable
  Distributed Systems (SRDS)}, pp.~253--255, IEEE, 2017.

\bibitem{gorenflo2020fastfabric}
C.~Gorenflo, S.~Lee, L.~Golab, and S.~Keshav, ``Fastfabric: Scaling hyperledger
  fabric to 20 000 transactions per second,'' {\em International Journal of
  Network Management}, vol.~30, no.~5, p.~e2099, 2020.

\bibitem{TANWAR2020102407}
S.~Tanwar, K.~Parekh, and R.~Evans, ``Blockchain-based electronic healthcare
  record system for healthcare 4.0 applications,'' {\em Journal of Information
  Security and Applications}, vol.~50, p.~102407, 2020.

\bibitem{dziembowski2018fairswap}
S.~Dziembowski, L.~Eckey, and S.~Faust, ``Fairswap: How to fairly exchange
  digital goods,'' in {\em Proceedings of the 2018 ACM SIGSAC Conference on
  Computer and Communications Security}, pp.~967--984, 2018.

\bibitem{tanwar2020blockchain}
S.~Tanwar, K.~Parekh, and R.~Evans, ``Blockchain-based electronic healthcare
  record system for healthcare 4.0 applications,'' {\em Journal of Information
  Security and Applications}, vol.~50, p.~102407, 2020.

\end{thebibliography}

	\newpage
	\onecolumn
	\appendices
    \section{Activity Sequence Diagram}
	
	let us take a close look at the activity/functional sequence diagrams that provide a transparent view of our system model. These schematic diagrams are the implementation view of the same system model, which was discussed in the earlier Section 3 as a high-level view. It specifies which entity will invoke which function of the smart contracts along the timeline. 

    \begin{figure}[H]
    \includegraphics[scale=0.5,angle=90]{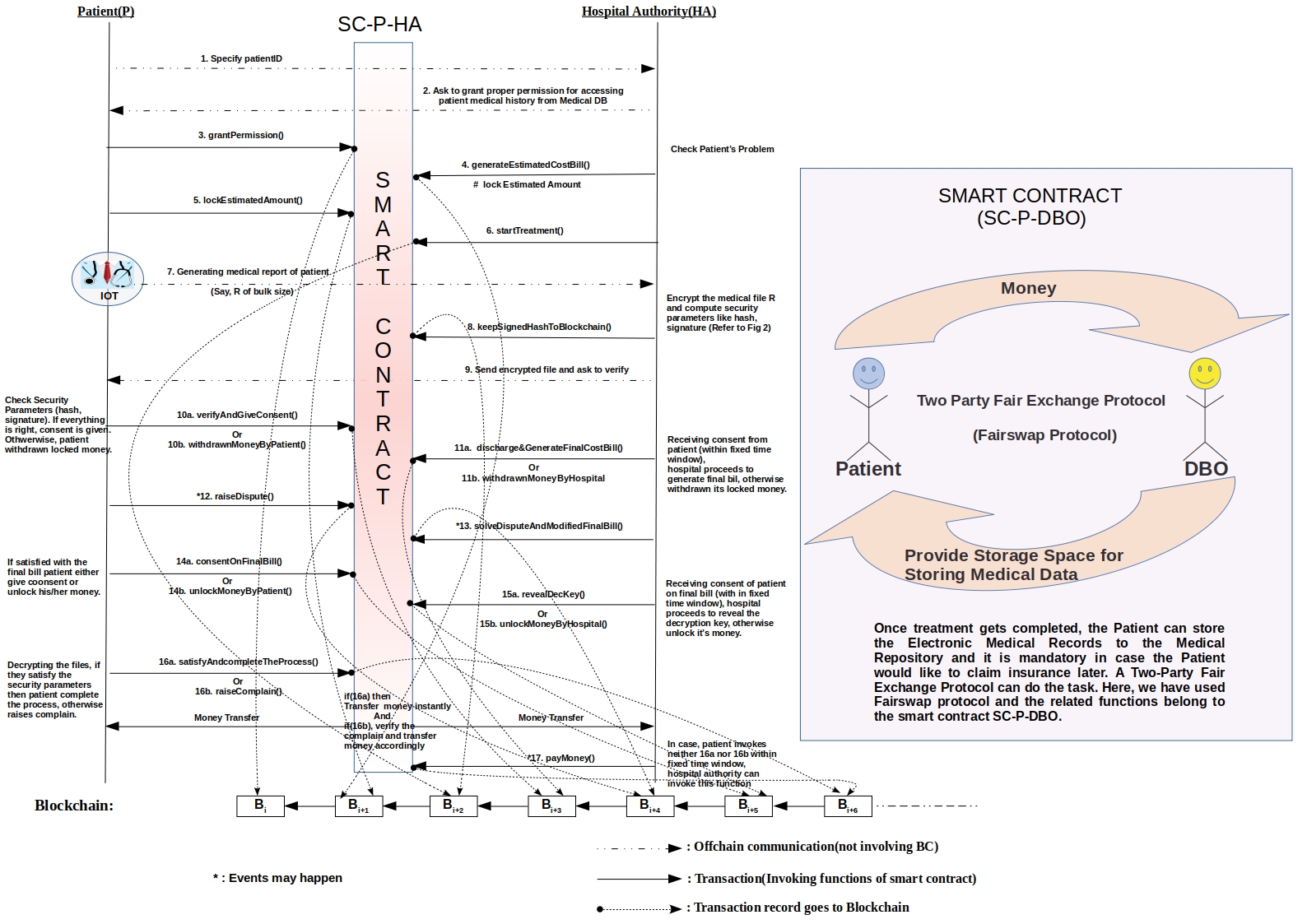}
    \centering
    \caption{Activity Sequence Diagram between Patient Hospital Authority and Database Owner}
    \end{figure}

    \begin{figure}[H]
    \includegraphics[scale=0.55,angle=90]{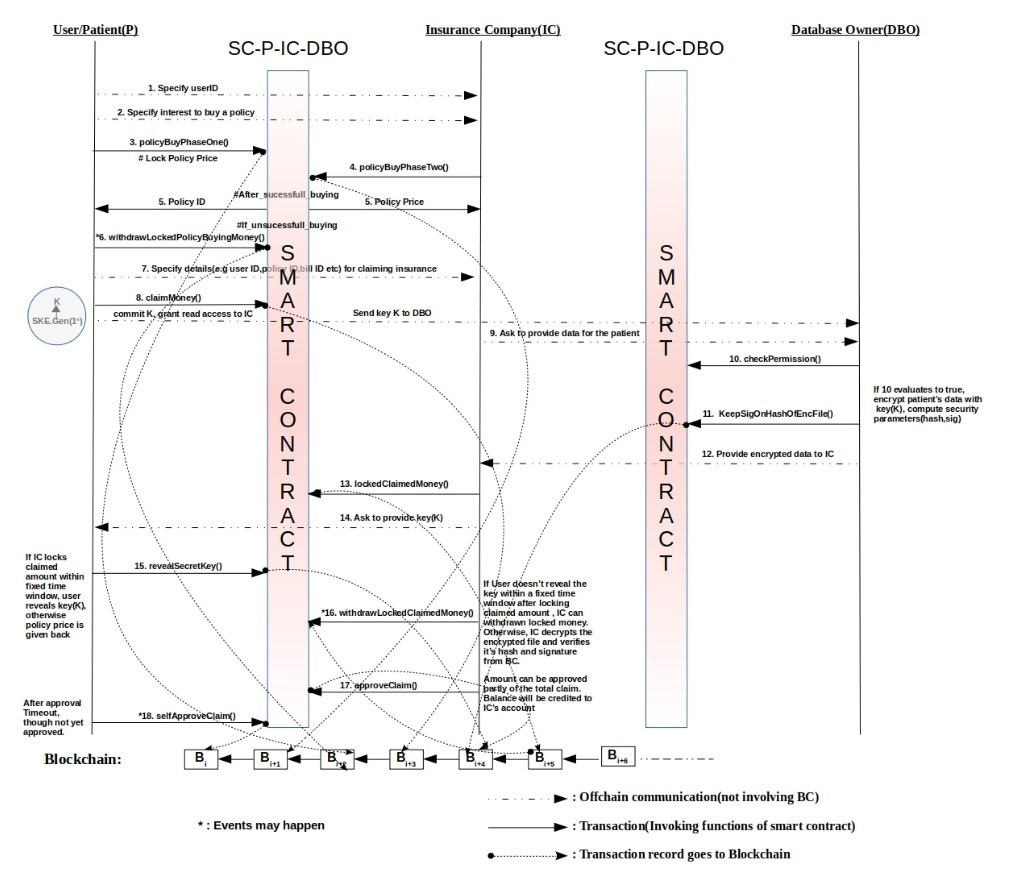}
    \centering
    \caption{Activity Sequence Diagram between User, Insurance Company and Database Owner}
    \end{figure}

    \begin{figure}[H]
    \includegraphics[scale=0.5,angle=90]{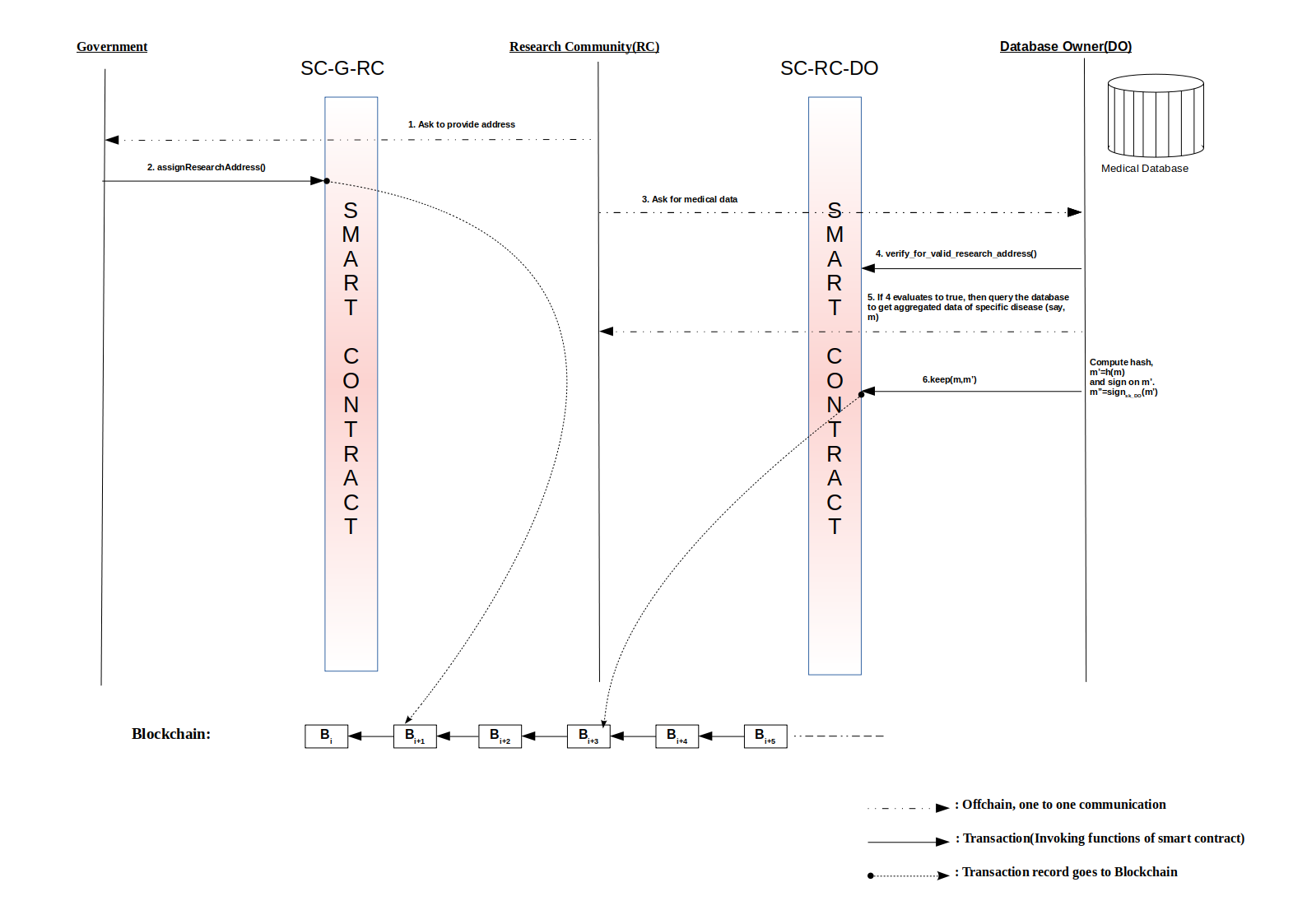}
    \centering
    \caption{Activity Sequence Diagram between Research Community and Database Owner}
    \end{figure}

	\newpage
	\section{Data Structures And Programming Constructs}
	
	Besides the primitives data types available in Solidity language (e.g., uint, address, bytes, bytes32, string, bool, etc.), we have also defined user-defined data types. These are mentioned below.

    1. \textbf{Structure:} To understand the algorithms clearly, we have to perceive the following major structure definitions that are used in our system implementation.
    
    \begin{figure}[H]
    \includegraphics[scale=0.6]{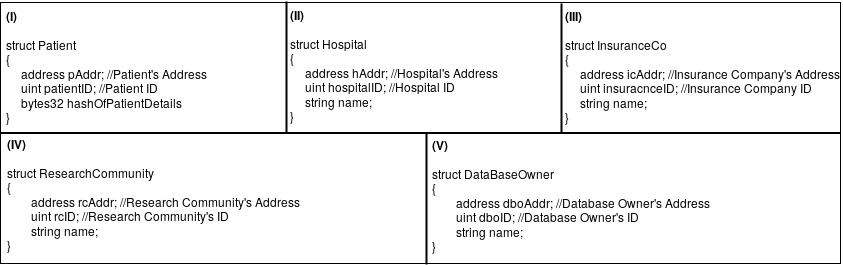}
    \centering
    \caption{Structure of Different Actors in the System(defined in SC\_Registration)}
    \end{figure}
    
    \begin{figure}[H]
    \includegraphics[scale=0.6]{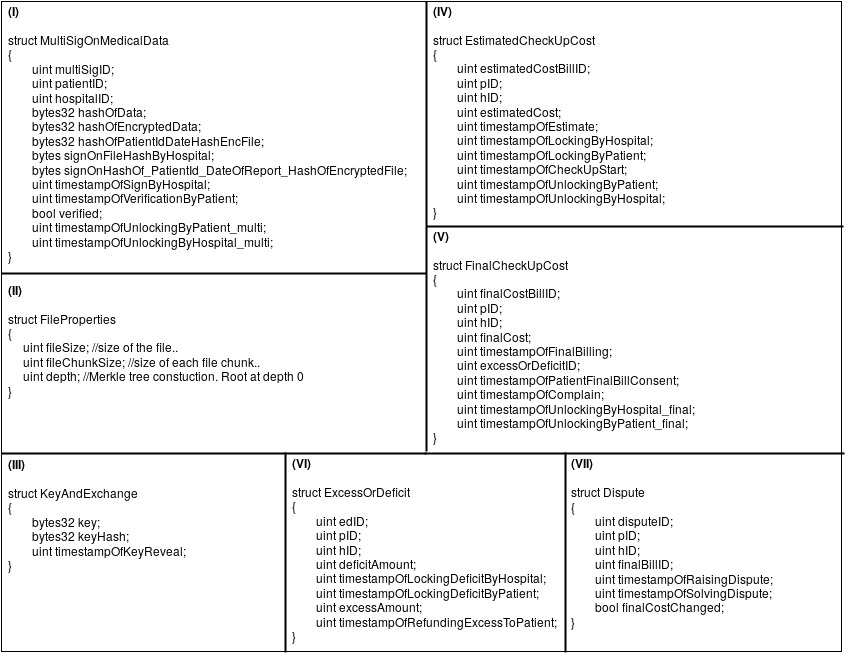}
    \centering
    \caption{Important Structures used in the Smart Contract SC\_P\_HA}
    \end{figure}
    
     \begin{figure}[H]
    \includegraphics[scale=0.6]{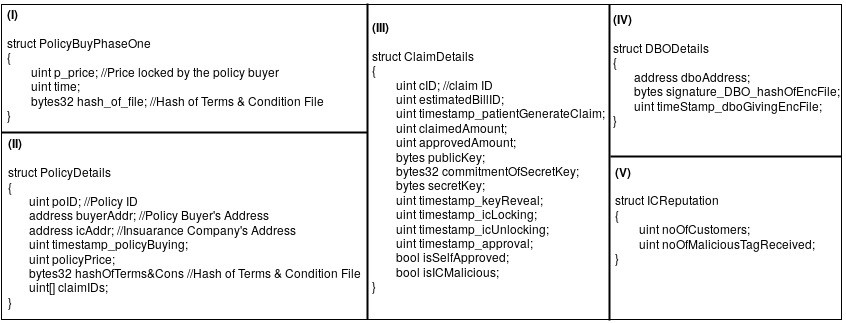}
    \centering
    \caption{Important Structures used in the Smart Contract SC\_P\_IC\_DBO}
    \end{figure}
    
    \begin{figure}[H]
    \includegraphics[scale=0.6]{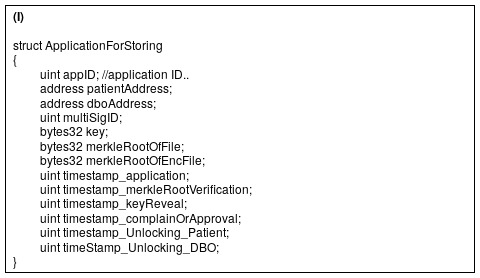}
    \centering
    \caption{Important Structures used in the Smart Contract SC\_P\_DBO}
    \label{fig:Image25}
    \end{figure}
    
    \begin{figure}[H]
    \includegraphics[scale=0.6]{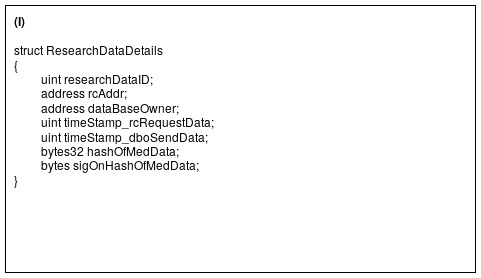}
    \centering
    \caption{Important Structures used in the Smart Contract SC\_RC\_DBO}
    \end{figure}
    
    \newpage
    
    2. \textbf{Array: }
    \newline
    \begin{table}[H]
        \centering
        \scalebox{0.8}
        {
            \begin{tabular}{| m{0.55cm} | m{3.62cm} | m{3.9cm} | m{8.7cm} |}
                \hline
                 \textbf{No} & \textbf{Array Name} & \textbf{Datatype} &\textbf{Description}  \\
                 \hline
                 \hline
                 1 &
                 patient[] &
                 struct Patient  &
                 When a new patient registers in the system, we make an entry into this array. The variable \textbf{PatientIDGenerator} keeps the count of the total no of registered patients/users in the system (or alternatively it specifies the length of the array). \\
                 \hline
                 2 &
                 hospital[] & 
                 struct Hospital & 
                 When a new hospital authority wants to register its name in the system, we make an entry into this array. The variable \textbf{HospitalIDGenerator} keeps the count of the total no of the registered hospitals in the system. \\
                 \hline
                 3 &
                 insuranceCo[] &
                 struct InsuranceCo &
                 When a new insurance company wants to register its name in the system, we make an entry into this array. The variable \textbf{insuranceCoIDGenerator} keeps the count of the total no of the registered insurance companies in the system. \\
                 \hline
                 4 &
                 researchCommunity[] &
                 struct ResearchCommunity  &
                 When a new research Community gets registered in the system by the Government, we make an entry into this array. The variable \textbf{rcIDGenerator} keeps the count of the total no of registered research communities in the system. \\
                 \hline
                 5 &
                 dataBaseOwner[] &
                 struct DataBaseOwner &
                 When a new database owner gets registered in the system by the Government, we make an entry into this array. The variable \textbf{dboIDGenerator} keeps the count of total no of registered DBO in the system. \\
                 \hline
                 6 &
                 estimatedCheckUpCost[] &
                 struct EstimatedCheckUpCost  & 
                 When a patient comes to a hospital for treatment, hospital generates the estimated cost bill. We make an entry of the same into this array. The variable \textbf{estimatedBillIDGenerator} keeps the count of total no of estimated bill generated in the system. \\
                 \hline
                 7 &
                 finalCheckUpCost[] &
                 struct FinalCheckUpCost & 
                 When a patient is about to be discharged, the hospital generates the final cost bill. We make an entry of the same into this array. The variable \textbf{finalBillIDGenerator} keeps the count of total no of the final bill generated in the system.\\
                 \hline
                 8 &
                 excessOrDeficit[] &
                 struct ExcessOrDeficit  &
                 When the final cost of treatment varies from the initial estimated cost, the hospital needs to instantiate a structure of ExcessOrDeficit. We make an entry of the same into this array. The variable \textbf{excessOrDeficitIDGenerator} keeps the count of total no of such instances in the system. \\ 
                 \hline
                 9 &
                 multiSigOnMedicalData[] &
                 struct MultiSigOnMedicalData & 
                 When a hospital is going to transfer the encrypted medical file to the patient, it instantiates a structure of  MultiSigOnMedicalData. The security-related aspects (e.g., hash, signature, timestamp, etc.) are taken care of by this structure. We make an entry of the same into this array. The variable \textbf{multiSigIDGenerator} keeps the count of the total no of such instances created in the system.
                 \\
                 \hline
                 10 &
                 applicationForStoring[] &
                 struct ApplicationForStoring  &
                 When a patient wants to store his/her medical files in the medical repository,(s)he applies to the DBO for the same. An entry is made in this array. A unique appID is assigned to each of these entries. The variable \textbf{applicationIDGenerator} keeps the count of the total no of such applications created in the system.\\
                 \hline
                 11 &
                 policyDetails[] &
                 struct PolicyDetails &
                 When a user buys a policy from an insurance company, an entry is made in this array. The variable \textbf{policyIDGenerator} helps to assign a unique policy ID against each such entry. \\
                 \hline
                 12 &
                 claimDetails[] &
                 struct ClaimDetails &
                 When a user claims for reimbursement of his/her medical bills against his policy, an entry is made in this array. The variable \textbf{claimIDGenerator} helps to assign a unique claim ID against each such entry. \\
                 \hline
            \end{tabular}
        }
        \caption{Major Arrays used in the Smart Contracts}
        \label{Major Arrays used in the Smart Contracts}
    \end{table}
    
    3. \textbf{Mapping:}
    \newline
    \begin{table}[H]
        \centering
        \scalebox{0.8}
        {
            \begin{tabular}{| m{0.55cm} | m{7.5cm} | m{8.5cm} |}
            \hline
                 \textbf{No} & \textbf{Mapping Name} & \textbf{Description} \\
                 \hline
                 \hline
                 1 &
                 mapFromAddrToID\_*\newline(* : P/H/IC/DBO/RC) &
                 Associating entity's address (Patient/ Hospital/ Insurance Company/ Database Owner/ Research community) to their corresponding ID. \\
            \hline
                2 &
                read\_permission\_to\_*\newline(* : H/IC) &
                Associating read access permission to the third party address (Hospital/ Insurance Company).\\
            \hline
                3 &
                patientID\_hospitalID\_estimatedBillID & 
                Coupling (patientID, hospitalID) to estimatedBillID.\\
            \hline
                4 &
                estimatedCostBillIDToMultiSigOnMedicalDataID &
                Mapping from Estimated Bill ID to corresponding Multi-Sig Data ID.\\
            \hline
                5 &
                estimatedBillIDToFinalBillID &
                Mapping from Estimated Cost Bill ID to corresponding Final Cost Bill ID.\\
            \hline
                6 &
                MultiSignIDToEncFileProperties &
                Refering to the Encrypted File properties(i.e struct FileProperty) from Multi-Sig Data ID.\\
            \hline
                7 &
                MultiSignIDToKeyAndExchange &
                Refering to the Encryption key properties(i.e struct KeyAndExchange) from Multi-Sig Data ID.\\
            \hline
                8 &
                moneyLocked &
                Coupling (buyer Address, IC Address) to struct policyBuyPhaseOne.\\
            \hline
                9 &
                reputationOfIC &
                Associating Insurance Company address to its reputation(i.e struct ICReputation).\\
            \hline
                10 &
                policyID\_map & Coupling (Policy buyer Address, Insurance Company address) to the policyID. \\
            \hline
                11 &
                securityMoney\_IC & Given an address of an IC, it refers to the amount of security money being locked in the system by the IC.\\
            \hline
                12 &
                deRegister\_IC &
                Given an address of an IC, it specifies whether the IC has been de-registered or not.\\
            \hline
                13 &
                cIDTopoID &
                Mapping from Claim ID to Policy ID.\\
            \hline
                14 &
                claimIDToDBODetails & Refering to the struct DBODetails from Claim ID.\\
            \hline
        
            \end{tabular}
        }
        \caption{Major Mappings used in the Smart Contracts}
        \label{Major Mappings used in the Smart Contracts}
    \end{table}
	
\end{document}